\documentclass[%
 reprint,
superscriptaddress,
groupedaddress,
 amsmath,amssymb,
 aps,
prl,
]{revtex4-2}

\usepackage{graphicx}
\usepackage{dcolumn}
\usepackage{bm}
\usepackage{comment}
\usepackage{hyperref}
\usepackage{xcolor}
\usepackage{slashed}
\usepackage[normalem]{ulem}

\newcommand \bea{\begin{eqnarray}}
\newcommand \eea{\end{eqnarray}}

\newcommand \qvec{{\bf q}}

\def\simge{\mathrel{%
       \rlap{\raise 0.511ex \hbox{$>$}}{\lower 0.511ex \hbox{$\sim$}}}}
\def\simle{\mathrel{
       \rlap{\raise 0.511ex \hbox{$<$}}{\lower 0.511ex \hbox{$\sim$}}}}
\def\beq{\begin{equation}}
\def\eeq{\end{equation}}
\def\w {\omega}

\definecolor{green}{rgb}{0.0,0.5, 0.0}

\begin{document}

\preprint{APS/123-QED}

\title{Electronic excitation spectra of molecular hydrogen in Phase I from Quantum Monte Carlo and Many-Body perturbation methods. }
\author{Vitaly Gorelov}
\affiliation{Laboratoire des Solides Irradiés, École Polytechnique, CNRS, CEA/DRF/IRAMIS, Institut Polytechnique de Paris, F-91128 Palaiseau, France.}
\affiliation{European Theoretical Spectroscopy Facility (ETSF)}
\author{Markus Holzmann}
\affiliation{Univ. Grenoble Alpes, CNRS, LPMMC, 38000 Grenoble, France}
\author{David M. Ceperley}
\affiliation{Department of Physics, University of Illinois Urbana-Champaign, Urbana, Illinois 61801, USA}
\author{Carlo Pierleoni}
\affiliation{Department of Physical and Chemical Sciences, University of L'Aquila, Via Vetoio 10, I-67010 L'Aquila, Italy}

\date{\today}

\begin{abstract}

We study the electronic excitation spectra in solid molecular hydrogen (phase I) at ambient temperature and 5-90 GPa pressures using Quantum Monte Carlo methods and Many-Body Perturbation Theory. In this range, the system changes from a wide gap molecular insulator to a semiconductor, altering the nature of the excitations from localized to delocalized. Computed gaps and spectra agree with experiments, proving the ability to predict accurately band gaps of many-body systems in presence of nuclear quantum and thermal effects. 

\end{abstract}
\maketitle

\maketitle



{\em Introduction} Many-body Hydrogen is a fundamental system whose physical properties have been the subject of numerous theoretical and experimental studies. Despite more than a century of investigations, its phase diagram under pressure is still uncertain because of experimental difficulties and computation inaccuracies\cite{Mao1994,McMahon2012a,Gregoryanz2020}. Of the many crystalline phases detected so far, only the crystalline structures of phase I, III and IV have been identified by X-ray diffraction \cite{Loubeyre1996,Ji2019,Ji2020}, while the structures of other phases have been predicted based on numerical algorithms \cite{Pickard2007,Pickard2012,Monserrat2018}. Similarly, characterization of the electronic properties, such as energy gaps and excitations, has been achieved mainly by optical probes, like absorption \cite{Loubeyre1996,Loubeyre2019} and reflection \cite{Dias2019} or by transport measurements \cite{Eremets2019}. In the search for metallic hydrogen, the electronic gap has been measured as a function of increasing pressure.
Recently, thanks to progress in high brilliance X-ray sources and in high-pressure experimental techniques\cite{Rueff2010,Shen2017,Mao2018}, inelastic X-ray scattering (IXS) has been successfully employed to detect the electronic excitation spectrum and extract the value of the electronic gap from the lower limit of the photon energy-loss spectra in Phase I \cite{Li2021}. 
From the theoretical perspective, the accurate calculation of 
optical properties and band gaps is difficult \cite{McMahon2012a}, 
since electron-phonon coupling as well as
excitonic effects are expected to play important roles.  

\begin{figure}
\center
\begin{minipage}[b]{\columnwidth}
\includegraphics[width=1.1\columnwidth]{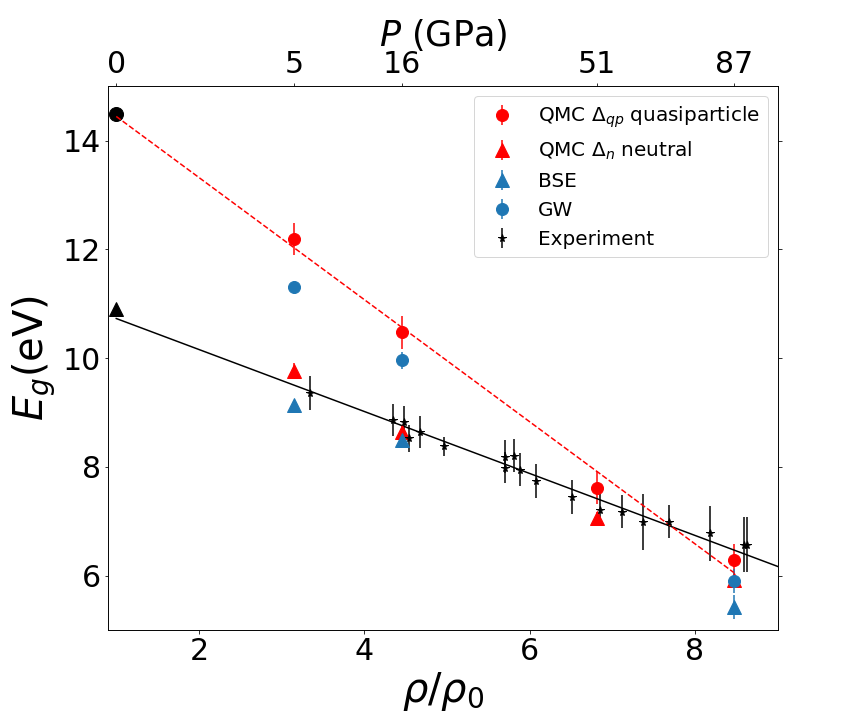}
\end{minipage}
\caption{\label{fig:gap} \small{Comparison between room temperature experimental data of ref. \cite{Li2021} and theoretical predictions for the electronic gap of solid hydrogen in phase I as a function of compression. We report quasi-particle (circles) and neutral gap from QMC (triangles) (red symbols) and from MBPT (blue symbols, triangles BSE, circles GW) both corrected for finite size effects. The black triangle and circle corresponds to the first exciton transition and the inter-band gap extracted from experimental absorption spectra at zero pressure $\rho/\rho_0=1$ 
The difference between the quasi-particle and neutral gap
is the exciton binding energy. 
The solid black line is a fit to experimental data; 
the red dashed line to QMC-QP gaps. 
}
}
\end{figure}

Here we present a detailed theoretical ab-initio study of the electronic excitation (absorption) spectra of Phase I hydrogen based
on Quantum Monte Carlo (QMC) and Many-Body perturbation theory
(MBPT) methods \cite{Martin2016}. Quantum and thermal effects of the protons are
included using Path Integral Monte Carlo calculations within the
Born-Oppenheimer approximation. Whereas the QMC calculations
focus on the value of the minimum excitation gap, we also
compute the energy-loss spectra based on the Bethe-Salpeter equation (BSE) to directly compare to experimental measurements.
Our calculations show that quantum nuclear effects reduce the
gap by $\sim 2$eV, a decrease only weakly dependant on pressure, in contrast to excitonic
effects which decrease more rapidly with pressure from $\sim 2$eV at three-fold compression to $\sim 0.5$eV at $90$GPa ($\sim$ nine-fold compression). 
Overall agreement, reported in Fig.\ref{fig:gap}, is observed between the QMC and BSE calculations
and experiment. 
The remaining small deviations with respect to the experimental values can be attributed to 
the extrapolation procedure, in particular the background subtraction
used to determine the energy gap from the experimental spectra.

Our results clearly point out the
limitations of self-consistent single electron theories like DFT.
Although
Ref.~\cite{Li2021} reports
DFT gaps with the HSE functional in agreement with experimental values, those calculations, based on ideal
crystal structures, rely on large 
error cancellations between the quantum
nuclear effects and the systematic 
underestimation of bandgaps 
of the DFT functional underlying the calculations.
(see the supplementary material of \cite{Li2021}). In addition, those calculations do not predict the changes between hydrogen and deuterium and
the strong pressure dependence of excitonic effects.

Previous MBPT \cite{Dvorak2014,Lebegue2012,Azadi2022,Kioupakis2008}
  or QMC \cite{Gorelov2020,Gorelov2020a,Azadi2017,Monacelli2023} studies of excitation gaps or optical properties have mostly focused on the high pressure regime
close to metallization. Since direct experimental results on structural properties are lacking in this region, comparison with experimental spectra \cite{Loubeyre2019} are less conclusive. Further, from a theoretical point of view, most of the studies are
not fully satisfying; Refs~\cite{Dvorak2014,Lebegue2012,Azadi2022,Kioupakis2008,Azadi2017} completely neglect quantum nuclear motion
whereas Ref.~\cite{Monacelli2023} is based on
QMC energies for the ideal structures augmented by DFT calculations for phonons
in the
self-consistent harmonic approximation and  electronic excitation spectra using different functionals.

{\em Methods.} Phase I of hydrogen has molecular centers on an HCP lattice with molecular orientations nearly isotropic. 
This phase is well characterized by X-ray diffraction at room temperature up to 120GPa \cite{Loubeyre1996}. A recent investigation extended the pressure range to phase III and phase IV up to 254GPa \cite{Ji2019} also providing the equation of state (EOS) and the cell geometry.

For our numerical study, we consider hydrogen molecules in the $P6_3/m$ structure with four molecules per unit cell. As in previous studies of hydrogen \cite{Gorelov2020a,Niu2023}, we employed a supercell with 48 molecules  ($N=96$ protons) comprising $3\times2\times2$ conventional cells (orthorhombic), a workable compromise between supercells with nearly cubic shape and a modest number of atoms. Molecules in the supercell were randomly oriented corresponding to the situation of phase I at room temperature \cite{Niu2023}.
We performed structural optimization of the molecular positions and supercell geometry at constant stress using the vdW-DF1 functional within DFT. This functional is among the best functionals for high pressure molecular hydrogen as benchmarked against QMC predictions\cite{Clay2014,Gorelov2019}. 
After geometry optimization we performed a room temperature NVT-Smart Monte Carlo simulation with both classical and quantum protons employing energies and forces from the DFT-vdW-DF1 functional to generate a set of uncorrelated configurations. 

This procedure was repeated at four different densities corresponding to compression values $\rho/\rho_0=3.15,~4.47,~6.86,~8.48$ ($r_s=2.21,~1.97,~1.71,~1.59$ respectively)  in order to investigate the pressure range between 5GPa and 90GPa. Here $\rho_0=0.0396$ $g/cm^3$  is the reference density at ambient pressure and cryogenic temperature.
Since the molecular geometry using DFT-vdW-DF1 are found to be accurate \cite{Clay2014,Gorelov2019}, we did not use the more expensive CEIMC algorithm (which relies on the QMC energies) for optimization.
Details of the thermodynamics and structures are reported in the Supplemental Material \cite{SM} (see also references \cite{Loubeyre1996,
Holzmann2003,
Pierleoni2008,
QE2009,
QE2017,
Morales2013,
Pierleoni2016,
Williams1951,
Lax1952,
Constable1975,
Lubos21,
Yang2020,
Li2021,
Ceperley1987,
Onida2002,
Strinati1988,
Hedin1965,
Adler1962,
Wiser1963,
Hybertsen1986,
Godby1988,
Albrecht1998,
Benedict1998,
Rohlfing2000,
Onida2002,
Gonze2005,
EXCcode,
Falco2013,
Gorelov2023_v2o5} therein). 

At each density and for each different system, we selected twenty independent configurations for the calculations of the electronic
excitations within QMC ten of which are also employed in the BSE calculations. Electronic energies are first
averaged over the nuclear configurations and excitation gaps
are obtained from the difference of averaged energies (see  the Supplemental Material \cite{SM} for details). 
Such a quantum average procedure becomes exact at low temperatures where zero point motion dominates the nuclear
trajectories as is the case for hydrogen at $T=300$K \cite{Gorelov2020,Gorelov2020b}.

For each nuclear configuration,
we first computed the fundamental or quasi-particle gap
\begin{equation}
\Delta_{qp} =E_0(N_e+1)+E_0(N_e-1)-2 E_0(N_e)
\label{eq:DeltaQP}
\end{equation}
adding and removing up to 6 electrons using reptation QMC
with a uniform positive background charge to have charge neutrality in the supercell.
To account for finite size effects, we have used grand canonical
twist averaging (GCTABC) and corrected for the leading order size
effects according to $\Delta_{qp}^{\infty}-\Delta_{qp}^L=|v_M(L)|/\epsilon$ as described in Ref.~\cite{Yang2020}.
Here, 
$v_M(L) \sim 1/L$ the Madelung constant (reported in the table in the Supplemental
  Material \cite{SM}), and 
$L$ the extension of the nearly cubic supercell. 
Heuristically, this $1/L$ dependence of the quasi-particle gap can be attributed to the additional charge interactions of the doped systems \cite{Makov1995,Engel1995}.
For all QMC calculations, the dielectric constant, $\epsilon$, used for size corrections has been extracted from extrapolating the long range behavior of the structure factor
(see the Supplemental Material \cite{SM}).

As a second step, we have also computed the neutral electron-hole
gap
\begin{equation}
\Delta_n = E_1(N_e)-E_0(N_e)
\label{eq:DeltaN}
\end{equation}
where $E_0(N_e)$ and $E_1(N_e)$ indicate electronic 
ground and first excited energies  
with $N_e$ electrons,
respectively. 
In practice, 
$\Delta_n$ is obtained within RQMC by promoting a single Bloch orbital from the ground state to an excited state in the Slater
determinant of the trial wave function 
\cite{Gorelov2023}. Kohn-Sham DFT energies are used
to determine the ordering. 

Accounting for the finite size effects of neutral excitations is more
delicate. For a fixed number of electrons,
the $1/L$ dependence will be absent for neutral excitations for a sufficiently large supercell since an
electron and a hole will be bound together forming a neutral object.
In practice, an apparent
$1/L$ behavior is still observed
\cite{Hunt2018,Gorelov2023} in situations where the
electron-hole attraction is not sufficiently strong so that the size of the exciton is larger or comparable with the size of the supercell. 
In order to quantitatively correct for finite-size effects,
additional information about the extension of electron-hole
pairs is needed.
An estimate of the excitonic length scale is $l_X=\epsilon/\mu$ where $\mu$ is the band mass 
describing the (extended) electron-hole excitation around the minimal gap (see the Supplemental Material \cite{SM}). Leading order size effects of neutral excitations are then estimated as \cite{Gorelov2023}
\begin{equation}
   \Delta_n^\infty-\Delta_n^L=\max \left[\frac{|v_M(L)|}{\epsilon}-\frac{|v_M(2l_X)|}{\epsilon},0 \right]
\label{eq:DeltaN_fse}
\end{equation}


For a subset of the configurations described above, we performed MBTP calculations on top of the DFT-LDA band structure. We employed both the GW and the BSE approach to compute the excitation spectra averaged over 10 configurations including both temperature and nuclear quantum effects. 
Whereas GW addresses quasi-particle excitations, BSE computes e-h spectra, including excitonic effects.

\begin{figure}
\center
\begin{minipage}[b]{1.\columnwidth}
\includegraphics[width=1.\columnwidth]{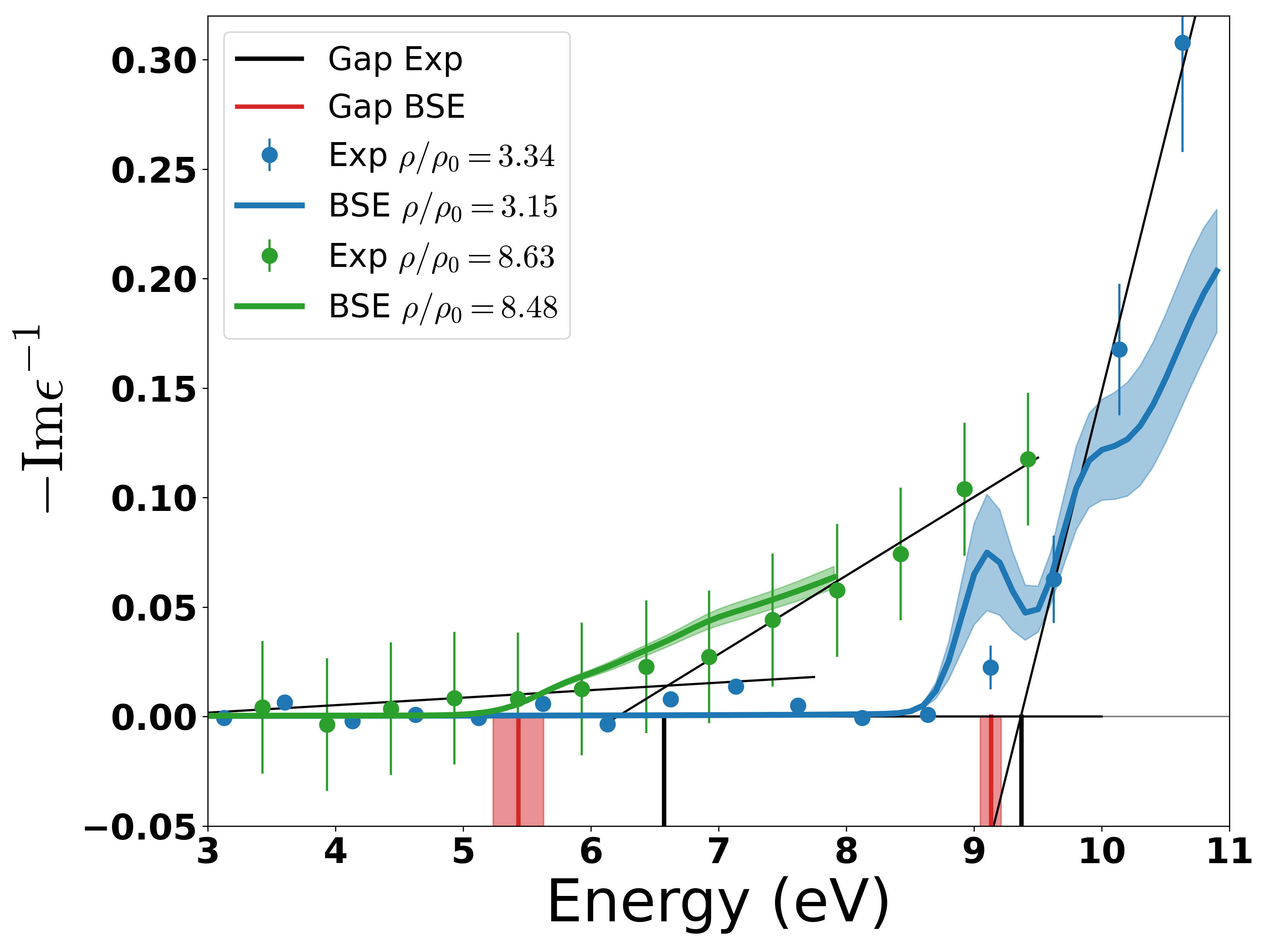}
\end{minipage}
\caption{\label{fig:spectra} Comparison of the measured and calculated (BSE) IXS spectra for the lowest (blue) and highest (green) calculated compressions. Closed circles with error bars are experimental data. Straight black lines are fit to the experimental data. Vertical black lines indicate the band gap extracted from the crossing of the fits at the two compressions. The vertical  red lines correspond to the BSE neutral gap for the corresponding compression. Only converged parts of the BSE spectra are shown.}
\end{figure}

To have a direct comparison between the QMC and BSE, we have performed the BSE calculations at vanishing momentum transfer. The IXS experiment measures the dynamic structure factor $S(\qvec,\w)=-q^2/(4\pi^2n)\text{Im}\epsilon_M^{-1}(\qvec,\w)$, where $n$ is the average electron density and $\epsilon_M(\qvec,\w)$ is a macroscopic dielectric function which can be directly computed within BSE. In our comparison to the IXS spectra, we examine the loss function at vanishing momentum: $-\lim_{\qvec \to 0}\text{Im}\epsilon_M^{-1}(\qvec,\w)$. Since the excitons in solid molecular hydrogen have a Frenkel - like nature with very little dispersion\cite{Cudazzo2013}, $\qvec \to 0$ is a good approximation of the spectral onset at finite momentum where the experiment is conducted. Note that the intensities in experimental IXS spectra are arbitrary. (see the Supplemental
  Material \cite{SM} for theoretical and computational details).
Values of the optical gap and the transition matrix elements computed using BSE were averaged over nuclear configurations, 
to obtain the spectra shown in  Fig. \ref{fig:spectra}.  A 0.2 eV Gaussian broadening was applied to the final averaged spectra.

\begin{figure}
\center
\begin{minipage}[b]{\columnwidth}
\includegraphics[width=1.\columnwidth]{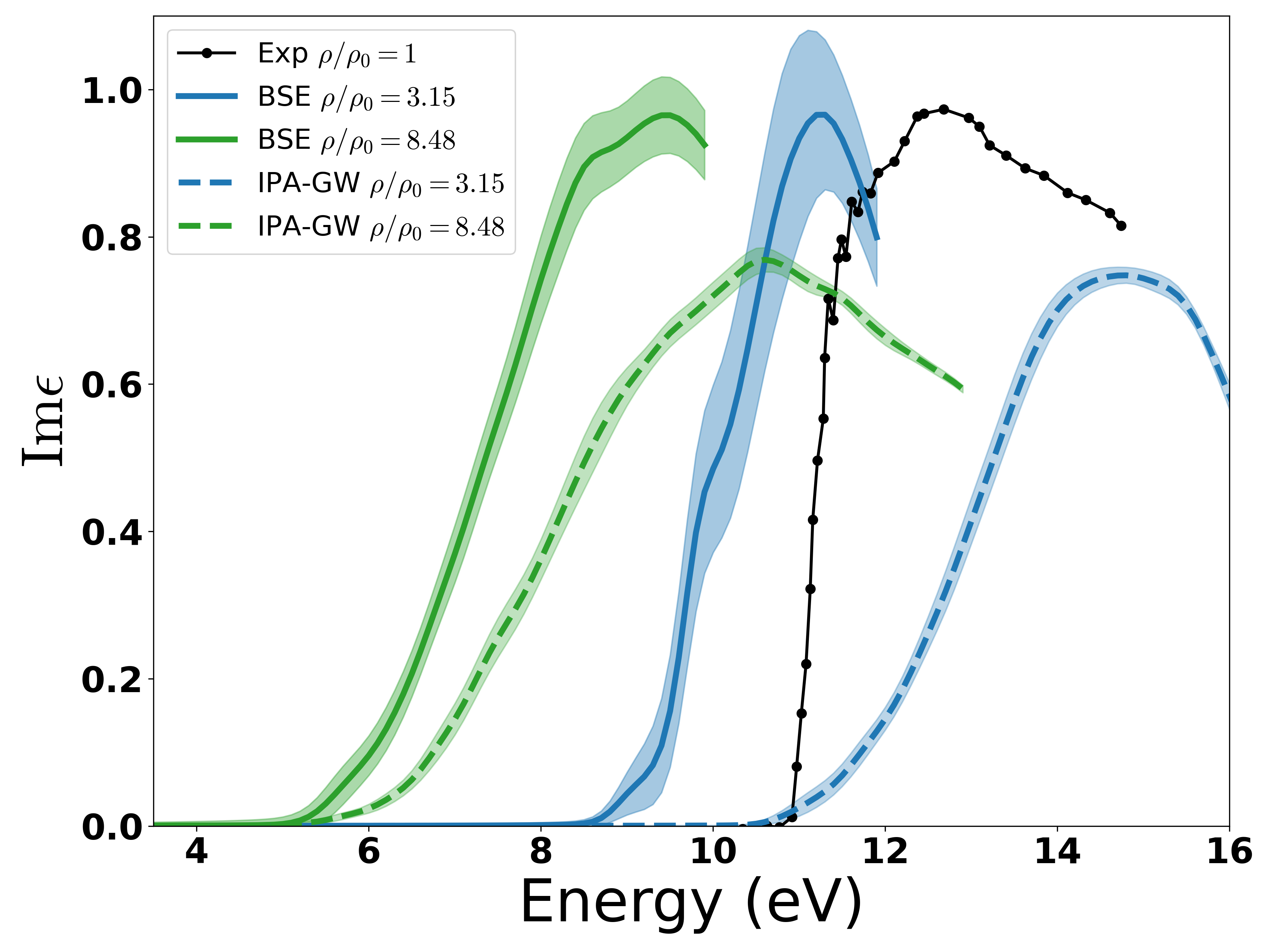}
\end{minipage}
\caption{\label{fig:abs} Absorption spectra from BSE (solid) and IPA-GW (dashed) at 
$\rho/\rho_0=8.48$ (green) and $\rho/\rho_0=3.15$ (blue) and experimental spectra at $\rho/\rho_0=1$(black) from \cite{Inoue1979}. We have renormalized the spectra to match the experimental intensity.} 
\end{figure}

\begin{figure}
\center
\begin{minipage}[b]{\columnwidth}
\includegraphics[width=1.\columnwidth]{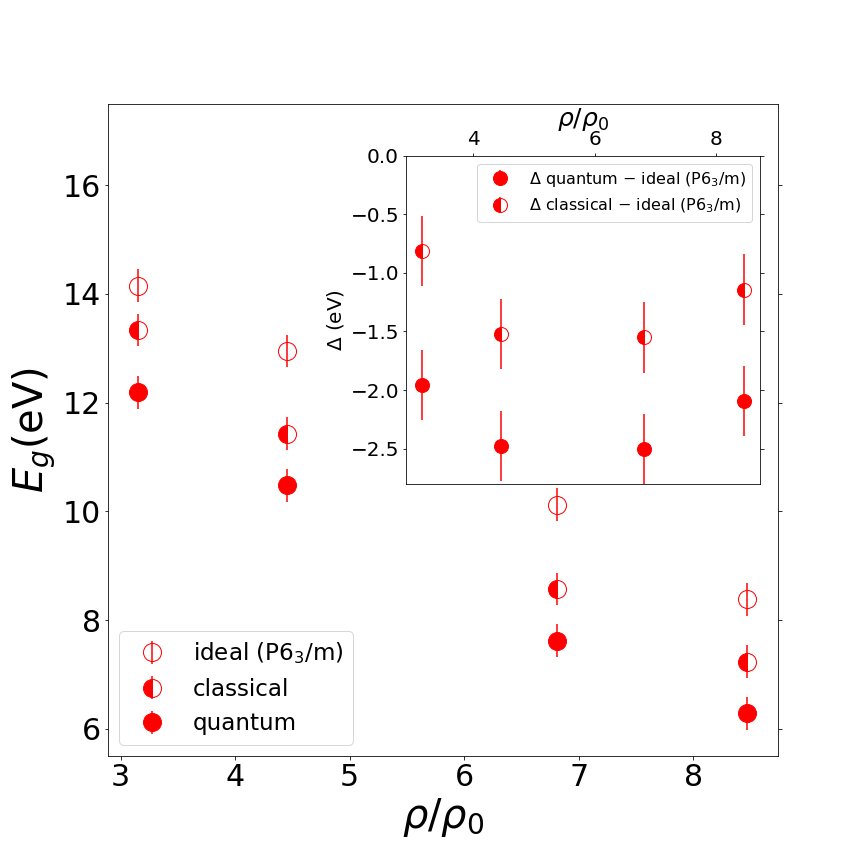}
\end{minipage}
\caption{\label{fig:nqe} Quasiparticle (QP) gap of the ideal $P6_3/m$ structure (open circles), QP gap with classical protons at room temperature (half circle) and QP gap with quantum protons at room temperature (filled circle).  Inset: The reduction of the quasiparticle gap due to temperature and quantum nuclear effects (filled circles) and with only temperature effects (half-filled circles)}
\end{figure}


{\em Results.} We have computed quasi-particle and neutral gaps in a compression range between three-fold to nine-fold using both QMC and MBPT
methods. 
Figure \ref{fig:gap} reports our results and compares with the experimental data of ref. \cite{Li2021}. 
Over the whole range of compression the system remains in the insulating state but the character of the
neutral excitation changes from localized to delocalized. 
Quasiparticle and neutral gaps from QMC are slightly larger than the GW 
and BSE results, respectively.
For the exciton binding energies $E_b$, (defined as the
difference between the neutral/BSE and the quasi-particle/GW gaps),
the agreement between QMC and MBPT is much better. See table I in the the Supplemental Material \cite{SM}.
A measure of localization of e-h pairs 
is the increase in the exciton binding energy
from $0.5$eV at nine-fold compression to $2$eV at our
lowest compression.


In general, we expect the quasi-particle/GW gap to describe the
onset of the continuum formed by inter-band transitions.  
A linear fit to the quasi-particle gap values extrapolates very close to the inter-band gap of 14.5 eV experimentally determined 
at zero pressure and cryogenic temperature ($\rho/\rho_0=1$) \cite{Inoue1979,Loubeyre2002}, and to the ionisation energy of hydrogen molecule (15eV) \cite{Beutler1936}. 
The neutral/BSE gap, instead, extrapolates to the first exciton transition measured by absorption at zero pressure \cite{Inoue1979}, which is only $\sim$0.2 eV lower than the free molecule excitation \cite{Herzberg1950}. 

The shape of the theoretical absorption spectra from BSE shown in Figure \ref{fig:abs} at our lowest pressure smoothly approaches the one measured in \cite{Inoue1979} at zero pressure.
Although the overall spectral structure seems to be preserved,
our analysis (see Fig. 7 of the Supplemental Material \cite{SM}) shows that
at our lowest density, the inter-band (GW) transitions start above $11$eV so that
the observed lower onset of absorption in Fig. \ref{fig:abs}
is intrinsically connected
to excitonic effects described by BSE.
This suggests that the excitons at low compression are tightly bound\cite{Davydov1971}, strongly localized on an individual molecule, supporting the 
interpretation 
of \cite{Inoue1979} at zero pressure.
At higher pressure, 
the binding energy decreases and the excitation becomes delocalized approaching pure inter-band transitions. This picture is further supported by the plot of the exciton wave function at different compressions (see Fig. 14 of the the Supplemental Material \cite{SM}). \footnote{We have verified that the first exciton energies in absorption spectra are identical to those in the energy loss spectra}

 
In order to further support the interpretation of excitonic 
effects in terms of free molecular excitations at the lowest compression,
we have performed QMC calculations 
for the neutral gap  
employing localized Gaussian molecular orbitals centered at each molecular center.
For each configuration, the lowest gap value is obtained by considering the excitation localized on the molecule with the longest bond length, corresponding to what was found in the BSE calculations. 
Despite an overall offset in total energies,
the value of the average gap using Gaussian molecular orbitals matches the one from our neutral calculation employing Bloch orbitals at the lowest compression. 
At higher compressions, this agreement is lost showing that the simple Gaussian approximation cannot describe the delocalization of the electrons in an exciton which extends
over neighboring molecules.



Let us now turn to the comparison of our neutral QMC and BSE gaps with the experimental values extracted
from the energy-loss edge of the IXS intensity
of Ref.~\cite{Li2021} shown in Fig.~\ref{fig:gap}. 
We observe an excellent agreement of theory and experiment. Insight into the origin of residual deviations can be obtained by comparing the IXS and BSE spectra. 
In figure \ref{fig:spectra} we report the comparison of BSE $-\text{Im}\epsilon_M^{-1}(\qvec \to 0,\w)$ and IXS spectra at low and high compression, together with the values of the BSE and experimentally extracted gaps  indicated by vertical bars. 
 At lower pressure/compression, the onset of energy loss is quite sharp, 
 the BSE optical gap coincides with the observed onset in the
 experimental spectra. However, to eliminate background effects,
 the experimental gap value reported in Ref.~\cite{Li2021}
 and shown in Fig.~\ref{fig:gap} is obtained by a linear
 extrapolation using points at higher energies. 
 Note that deviations of BSE with respect to experimental data at higher
 energies are an artifact of the limited number of unoccupied bands
 taken into account in the BSE calculation.
Possible bias due to the linear extrapolation roughly coincides
with the experimental errors quoted at this compression which 
must be interpreted as a systematic error.
Within BSE, the onset of energy loss is due to the sharp and intense first excitonic peak 
whose intensity might be larger than in experiment due to the vanishing momentum transfer in the BSE calculations 
 (see Fig. 8 of the the Supplemental Material \cite{SM}).
 
At higher pressure, where the excitonic intensities are weaker (see Fig. 8 of the the Supplemental Material \cite{SM}), the onset is smeared out. 
The values of the experimental gap at higher
compressions have been estimated as the intersection point between
two slopes; the less steep one is attributed to residual beryllium gasket background effects.
(see the less steep fit of the right panel of Fig.~\ref{fig:spectra}). In our BSE calculations, however, we observe that the IXS spectrum begins at 5.3 eV with a very weak first excitonic peak such that the BSE
results are within the systematic uncertainty
of the experimental gap determination.

In order to quantify the influence of thermal and quantum nuclear effects on electronic gaps we computed the gap for systems of classical protons and for the ideal $P6_3/m$ structure at the four compression values. In Fig. (\ref{fig:nqe}) we report $\Delta_{qp}$ for the relaxed $P6_3/m$ structure, for systems of both classical and of quantum protons at 300K. Ideal structures have the largest gaps, thermal effects alone (classical protons) provide roughly $\sim$1eV reduction of the gap while nuclear quantum effects provide an additional $\sim$1eV reduction of the gap, roughly independent of compression (see the inset to Fig. \ref{fig:nqe}). We expect that the deuterium gap will be half way between the gap of hydrogen and that of classical protons. Note that we do not consider any effects of quantum statistics on the molecular rotational spectra.

{\em Conclusion.} The pressure induced variation of solid hydrogen
from a wide gap insulator towards a metal has been challenging experiment and theory for decades.  
We have made a theoretical study of the electronic excitation
gap and spectral properties based on QMC and MBPT methods of Phase I where quantitative
comparison to IXS measurements are possible. We have shown that
quantum nuclear and excitonic effects introduce sizeable
reductions of the gap. In contrast to thermal and
quantum nuclear effects, the reduction of the gap due to excitonic effects decreases rapidly with pressure. At our highest compression,
quasiparticle/GW and neutral/BSE gaps almost coincide.
Therefore, the roughly linear behavior of the
closing of the gap in the range of compressions studied here
and in Ref.~\cite{Li2021} will change slope around $100$GPa
to follow the line of the quasi-particle gap.

Our calculations put forward a pressure induced cross-over
of the optical excitation spectra from a typical molecular 
crystal towards a semiconductor-like behavior. At low pressure,
excitons are mainly localized on molecular centers and form
a broad excitonic band. At high compression, energy
loss and absorption spectra are dominated by quasi-particle
excitations
with weakly bound excitons, delocalized over several unit cells.
We have shown that nuclear quantum effects and intrinsic many-body calculations (MBPT or QMC)
are needed for a quantitative description.



{\em Acknowledgements.}  D. M. C. is supported by DOE DE-SC0020177. CP was supported by the European Union - NextGenerationEU under the Italian Ministry of University and Research (MUR) projects PRIN2022-2022NRBLPT CUP E53D23001790006 and PRIN2022-P2022MC742PNRR, CUP E53D23018440001.
This research used HPC resources from GENCI-IDRIS 2022-AD010912502R1, 2023-A0140914158 and GENCI (Project No. 544).

\bibliography{PhaseIgap}

\providecommand{\noopsort}[1]{}\providecommand{\singleletter}[1]{#1}%
\begin{thebibliography}{68}%
\makeatletter
\providecommand \@ifxundefined [1]{%
 \@ifx{#1\undefined}
}%
\providecommand \@ifnum [1]{%
 \ifnum #1\expandafter \@firstoftwo
 \else \expandafter \@secondoftwo
 \fi
}%
\providecommand \@ifx [1]{%
 \ifx #1\expandafter \@firstoftwo
 \else \expandafter \@secondoftwo
 \fi
}%
\providecommand \natexlab [1]{#1}%
\providecommand \enquote  [1]{``#1''}%
\providecommand \bibnamefont  [1]{#1}%
\providecommand \bibfnamefont [1]{#1}%
\providecommand \citenamefont [1]{#1}%
\providecommand \href@noop [0]{\@secondoftwo}%
\providecommand \href [0]{\begingroup \@sanitize@url \@href}%
\providecommand \@href[1]{\@@startlink{#1}\@@href}%
\providecommand \@@href[1]{\endgroup#1\@@endlink}%
\providecommand \@sanitize@url [0]{\catcode `\\12\catcode `\$12\catcode
  `\&12\catcode `\#12\catcode `\^12\catcode `\_12\catcode `\%12\relax}%
\providecommand \@@startlink[1]{}%
\providecommand \@@endlink[0]{}%
\providecommand \url  [0]{\begingroup\@sanitize@url \@url }%
\providecommand \@url [1]{\endgroup\@href {#1}{\urlprefix }}%
\providecommand \urlprefix  [0]{URL }%
\providecommand \Eprint [0]{\href }%
\providecommand \doibase [0]{https://doi.org/}%
\providecommand \selectlanguage [0]{\@gobble}%
\providecommand \bibinfo  [0]{\@secondoftwo}%
\providecommand \bibfield  [0]{\@secondoftwo}%
\providecommand \translation [1]{[#1]}%
\providecommand \BibitemOpen [0]{}%
\providecommand \bibitemStop [0]{}%
\providecommand \bibitemNoStop [0]{.\EOS\space}%
\providecommand \EOS [0]{\spacefactor3000\relax}%
\providecommand \BibitemShut  [1]{\csname bibitem#1\endcsname}%
\let\auto@bib@innerbib\@empty
\bibitem [{\citenamefont {Mao}\ and\ \citenamefont {Hemley}(1994)}]{Mao1994}%
  \BibitemOpen
  \bibfield  {author} {\bibinfo {author} {\bibfnamefont {H.~K.}\ \bibnamefont
  {Mao}}\ and\ \bibinfo {author} {\bibfnamefont {R.}~\bibnamefont {Hemley}},\
  }\bibfield  {title} {\bibinfo {title} {{Ultrahigh-pressure transitions in
  solid hydrogen}},\ }\href@noop {} {\bibfield  {journal} {\bibinfo  {journal}
  {Rev. Mod. Phys.}\ }\textbf {\bibinfo {volume} {66}},\ \bibinfo {pages} {671}
  (\bibinfo {year} {1994})}\BibitemShut {NoStop}%
\bibitem [{\citenamefont {McMahon}\ \emph {et~al.}(2012)\citenamefont
  {McMahon}, \citenamefont {Morales}, \citenamefont {Pierleoni},\ and\
  \citenamefont {Ceperley}}]{McMahon2012a}%
  \BibitemOpen
  \bibfield  {author} {\bibinfo {author} {\bibfnamefont {J.~M.}\ \bibnamefont
  {McMahon}}, \bibinfo {author} {\bibfnamefont {M.~A.}\ \bibnamefont
  {Morales}}, \bibinfo {author} {\bibfnamefont {C.}~\bibnamefont {Pierleoni}},\
  and\ \bibinfo {author} {\bibfnamefont {D.~M.}\ \bibnamefont {Ceperley}},\
  }\bibfield  {title} {\bibinfo {title} {{The properties of hydrogen and helium
  under extreme conditions}},\ }\href
  {https://doi.org/10.1103/RevModPhys.84.1607} {\bibfield  {journal} {\bibinfo
  {journal} {Rev. Mod. Phys.}\ }\textbf {\bibinfo {volume} {84}},\ \bibinfo
  {pages} {1607} (\bibinfo {year} {2012})}\BibitemShut {NoStop}%
\bibitem [{\citenamefont {Gregoryanz}\ \emph {et~al.}(2020)\citenamefont
  {Gregoryanz}, \citenamefont {Ji}, \citenamefont {Dalladay-Simpson},
  \citenamefont {Li}, \citenamefont {Howie},\ and\ \citenamefont
  {Mao}}]{Gregoryanz2020}%
  \BibitemOpen
  \bibfield  {author} {\bibinfo {author} {\bibfnamefont {E.}~\bibnamefont
  {Gregoryanz}}, \bibinfo {author} {\bibfnamefont {C.}~\bibnamefont {Ji}},
  \bibinfo {author} {\bibfnamefont {P.}~\bibnamefont {Dalladay-Simpson}},
  \bibinfo {author} {\bibfnamefont {B.}~\bibnamefont {Li}}, \bibinfo {author}
  {\bibfnamefont {R.~T.}\ \bibnamefont {Howie}},\ and\ \bibinfo {author}
  {\bibfnamefont {H.-K.}\ \bibnamefont {Mao}},\ }\bibfield  {title} {\bibinfo
  {title} {{Everything you always wanted to know about metallic hydrogen but
  were afraid to ask}},\ }\href {https://doi.org/10.1063/5.0002104} {\bibfield
  {journal} {\bibinfo  {journal} {Matter and Radiation at Extremes}\ }\textbf
  {\bibinfo {volume} {5}},\ \bibinfo {pages} {038101} (\bibinfo {year}
  {2020})}\BibitemShut {NoStop}%
\bibitem [{\citenamefont {Loubeyre}\ \emph {et~al.}(1996)\citenamefont
  {Loubeyre}, \citenamefont {LeToullec}, \citenamefont {Hausermann},
  \citenamefont {Hanfland}, \citenamefont {Hemley}, \citenamefont {Mao},\ and\
  \citenamefont {Finger}}]{Loubeyre1996}%
  \BibitemOpen
  \bibfield  {author} {\bibinfo {author} {\bibfnamefont {P.}~\bibnamefont
  {Loubeyre}}, \bibinfo {author} {\bibfnamefont {R.}~\bibnamefont {LeToullec}},
  \bibinfo {author} {\bibfnamefont {D.}~\bibnamefont {Hausermann}}, \bibinfo
  {author} {\bibfnamefont {M.}~\bibnamefont {Hanfland}}, \bibinfo {author}
  {\bibfnamefont {R.~J.}\ \bibnamefont {Hemley}}, \bibinfo {author}
  {\bibfnamefont {H.~K.}\ \bibnamefont {Mao}},\ and\ \bibinfo {author}
  {\bibfnamefont {L.~W.}\ \bibnamefont {Finger}},\ }\bibfield  {title}
  {\bibinfo {title} {{X-ray diffraction and equation of state of hydrogen at
  megabar pressures}},\ }\href@noop {} {\bibfield  {journal} {\bibinfo
  {journal} {Nature}\ }\textbf {\bibinfo {volume} {383}},\ \bibinfo {pages}
  {702} (\bibinfo {year} {1996})}\BibitemShut {NoStop}%
\bibitem [{\citenamefont {Ji}\ \emph {et~al.}(2019)\citenamefont {Ji},
  \citenamefont {Li}, \citenamefont {Liu}, \citenamefont {Smith}, \citenamefont
  {Majumdar}, \citenamefont {Luo}, \citenamefont {Ahuja}, \citenamefont {Shu},
  \citenamefont {Wang}, \citenamefont {Sinogeikin}, \citenamefont {Meng},
  \citenamefont {Prakapenka}, \citenamefont {Greenberg}, \citenamefont {Xu},
  \citenamefont {Huang}, \citenamefont {Yang}, \citenamefont {Shen},
  \citenamefont {Mao},\ and\ \citenamefont {Mao}}]{Ji2019}%
  \BibitemOpen
  \bibfield  {author} {\bibinfo {author} {\bibfnamefont {C.}~\bibnamefont
  {Ji}}, \bibinfo {author} {\bibfnamefont {B.}~\bibnamefont {Li}}, \bibinfo
  {author} {\bibfnamefont {W.}~\bibnamefont {Liu}}, \bibinfo {author}
  {\bibfnamefont {J.~S.}\ \bibnamefont {Smith}}, \bibinfo {author}
  {\bibfnamefont {A.}~\bibnamefont {Majumdar}}, \bibinfo {author}
  {\bibfnamefont {W.}~\bibnamefont {Luo}}, \bibinfo {author} {\bibfnamefont
  {R.}~\bibnamefont {Ahuja}}, \bibinfo {author} {\bibfnamefont
  {J.}~\bibnamefont {Shu}}, \bibinfo {author} {\bibfnamefont {J.}~\bibnamefont
  {Wang}}, \bibinfo {author} {\bibfnamefont {S.}~\bibnamefont {Sinogeikin}},
  \bibinfo {author} {\bibfnamefont {Y.}~\bibnamefont {Meng}}, \bibinfo {author}
  {\bibfnamefont {V.~B.}\ \bibnamefont {Prakapenka}}, \bibinfo {author}
  {\bibfnamefont {E.}~\bibnamefont {Greenberg}}, \bibinfo {author}
  {\bibfnamefont {R.}~\bibnamefont {Xu}}, \bibinfo {author} {\bibfnamefont
  {X.}~\bibnamefont {Huang}}, \bibinfo {author} {\bibfnamefont
  {W.}~\bibnamefont {Yang}}, \bibinfo {author} {\bibfnamefont {G.}~\bibnamefont
  {Shen}}, \bibinfo {author} {\bibfnamefont {W.~L.}\ \bibnamefont {Mao}},\ and\
  \bibinfo {author} {\bibfnamefont {H.-K.}\ \bibnamefont {Mao}},\ }\bibfield
  {title} {\bibinfo {title} {{Ultrahigh-pressure isostructural electronic
  transitions in hydrogen}},\ }\href
  {https://doi.org/10.1038/s41586-019-1565-9} {\bibfield  {journal} {\bibinfo
  {journal} {Nature}\ }\textbf {\bibinfo {volume} {573}},\ \bibinfo {pages}
  {558} (\bibinfo {year} {2019})}\BibitemShut {NoStop}%
\bibitem [{\citenamefont {Ji}\ \emph {et~al.}(2020)\citenamefont {Ji},
  \citenamefont {Li}, \citenamefont {Liu}, \citenamefont {Smith}, \citenamefont
  {Bj{\"{o}}rling}, \citenamefont {Majumdar}, \citenamefont {Luo},
  \citenamefont {Ahuja}, \citenamefont {Shu}, \citenamefont {Wang},
  \citenamefont {Sinogeikin}, \citenamefont {Meng}, \citenamefont {Prakapenka},
  \citenamefont {Greenberg}, \citenamefont {Xu}, \citenamefont {Huang},
  \citenamefont {Ding}, \citenamefont {Soldatov}, \citenamefont {Yang},
  \citenamefont {Shen}, \citenamefont {Mao},\ and\ \citenamefont
  {Mao}}]{Ji2020}%
  \BibitemOpen
  \bibfield  {author} {\bibinfo {author} {\bibfnamefont {C.}~\bibnamefont
  {Ji}}, \bibinfo {author} {\bibfnamefont {B.}~\bibnamefont {Li}}, \bibinfo
  {author} {\bibfnamefont {W.}~\bibnamefont {Liu}}, \bibinfo {author}
  {\bibfnamefont {J.~S.}\ \bibnamefont {Smith}}, \bibinfo {author}
  {\bibfnamefont {A.}~\bibnamefont {Bj{\"{o}}rling}}, \bibinfo {author}
  {\bibfnamefont {A.}~\bibnamefont {Majumdar}}, \bibinfo {author}
  {\bibfnamefont {W.}~\bibnamefont {Luo}}, \bibinfo {author} {\bibfnamefont
  {R.}~\bibnamefont {Ahuja}}, \bibinfo {author} {\bibfnamefont
  {J.}~\bibnamefont {Shu}}, \bibinfo {author} {\bibfnamefont {J.}~\bibnamefont
  {Wang}}, \bibinfo {author} {\bibfnamefont {S.}~\bibnamefont {Sinogeikin}},
  \bibinfo {author} {\bibfnamefont {Y.}~\bibnamefont {Meng}}, \bibinfo {author}
  {\bibfnamefont {V.~B.}\ \bibnamefont {Prakapenka}}, \bibinfo {author}
  {\bibfnamefont {E.}~\bibnamefont {Greenberg}}, \bibinfo {author}
  {\bibfnamefont {R.}~\bibnamefont {Xu}}, \bibinfo {author} {\bibfnamefont
  {X.}~\bibnamefont {Huang}}, \bibinfo {author} {\bibfnamefont
  {Y.}~\bibnamefont {Ding}}, \bibinfo {author} {\bibfnamefont {A.}~\bibnamefont
  {Soldatov}}, \bibinfo {author} {\bibfnamefont {W.}~\bibnamefont {Yang}},
  \bibinfo {author} {\bibfnamefont {G.}~\bibnamefont {Shen}}, \bibinfo {author}
  {\bibfnamefont {W.~L.}\ \bibnamefont {Mao}},\ and\ \bibinfo {author}
  {\bibfnamefont {H.-K.}\ \bibnamefont {Mao}},\ }\bibfield  {title} {\bibinfo
  {title} {{Crystallography of low Z material at ultrahigh pressure: Case study
  on solid hydrogen}},\ }\href {https://doi.org/10.1063/5.0003288} {\bibfield
  {journal} {\bibinfo  {journal} {Matter and Radiation at Extremes}\ }\textbf
  {\bibinfo {volume} {5}},\ \bibinfo {pages} {038401} (\bibinfo {year}
  {2020})}\BibitemShut {NoStop}%
\bibitem [{\citenamefont {Pickard}\ and\ \citenamefont
  {Needs}(2007)}]{Pickard2007}%
  \BibitemOpen
  \bibfield  {author} {\bibinfo {author} {\bibfnamefont {C.~J.}\ \bibnamefont
  {Pickard}}\ and\ \bibinfo {author} {\bibfnamefont {R.~J.}\ \bibnamefont
  {Needs}},\ }\bibfield  {title} {\bibinfo {title} {{Structure of phase III of
  solid hydrogen}},\ }\href {https://doi.org/10.1038/nphys625} {\bibfield
  {journal} {\bibinfo  {journal} {Nature Physics}\ }\textbf {\bibinfo {volume}
  {3}},\ \bibinfo {pages} {473} (\bibinfo {year} {2007})}\BibitemShut {NoStop}%
\bibitem [{\citenamefont {Pickard}\ \emph {et~al.}(2012)\citenamefont
  {Pickard}, \citenamefont {Martinez-Canales},\ and\ \citenamefont
  {Needs}}]{Pickard2012}%
  \BibitemOpen
  \bibfield  {author} {\bibinfo {author} {\bibfnamefont {C.~J.}\ \bibnamefont
  {Pickard}}, \bibinfo {author} {\bibfnamefont {M.}~\bibnamefont
  {Martinez-Canales}},\ and\ \bibinfo {author} {\bibfnamefont {R.~J.}\
  \bibnamefont {Needs}},\ }\bibfield  {title} {\bibinfo {title} {{Density
  functional theory study of phase IV of solid hydrogen}},\ }\href
  {https://doi.org/10.1103/PhysRevB.85.214114} {\bibfield  {journal} {\bibinfo
  {journal} {Phys. Rev. B}\ }\textbf {\bibinfo {volume} {85}},\ \bibinfo
  {pages} {214114} (\bibinfo {year} {2012})}\BibitemShut {NoStop}%
\bibitem [{\citenamefont {Monserrat}\ \emph {et~al.}(2018)\citenamefont
  {Monserrat}, \citenamefont {Drummond}, \citenamefont {Dalladay-simpson},
  \citenamefont {Howie}, \citenamefont {{Lopez Rios}}, \citenamefont
  {Gregoryanz}, \citenamefont {Pickard},\ and\ \citenamefont
  {Needs}}]{Monserrat2018}%
  \BibitemOpen
  \bibfield  {author} {\bibinfo {author} {\bibfnamefont {B.}~\bibnamefont
  {Monserrat}}, \bibinfo {author} {\bibfnamefont {N.~D.}\ \bibnamefont
  {Drummond}}, \bibinfo {author} {\bibfnamefont {P.}~\bibnamefont
  {Dalladay-simpson}}, \bibinfo {author} {\bibfnamefont {R.~T.}\ \bibnamefont
  {Howie}}, \bibinfo {author} {\bibfnamefont {P.}~\bibnamefont {{Lopez Rios}}},
  \bibinfo {author} {\bibfnamefont {E.}~\bibnamefont {Gregoryanz}}, \bibinfo
  {author} {\bibfnamefont {C.~J.}\ \bibnamefont {Pickard}},\ and\ \bibinfo
  {author} {\bibfnamefont {R.~J.}\ \bibnamefont {Needs}},\ }\bibfield  {title}
  {\bibinfo {title} {{Structure and metallicity of phase V of hydrogen}},\
  }\href@noop {} {\bibfield  {journal} {\bibinfo  {journal} {Phys Rev Letts}\
  }\textbf {\bibinfo {volume} {120}},\ \bibinfo {pages} {255701} (\bibinfo
  {year} {2018})}\BibitemShut {NoStop}%
\bibitem [{\citenamefont {Loubeyre}\ \emph {et~al.}(2020)\citenamefont
  {Loubeyre}, \citenamefont {Occelli},\ and\ \citenamefont
  {Dumas}}]{Loubeyre2019}%
  \BibitemOpen
  \bibfield  {author} {\bibinfo {author} {\bibfnamefont {P.}~\bibnamefont
  {Loubeyre}}, \bibinfo {author} {\bibfnamefont {F.}~\bibnamefont {Occelli}},\
  and\ \bibinfo {author} {\bibfnamefont {P.}~\bibnamefont {Dumas}},\ }\bibfield
   {title} {\bibinfo {title} {{Synchrotron infrared spectroscopic evidence of
  the probable transition to metal hydrogen}},\ }\href
  {https://doi.org/10.1038/s41586-019-1927-3} {\bibfield  {journal} {\bibinfo
  {journal} {Nature}\ }\textbf {\bibinfo {volume} {577}},\ \bibinfo {pages}
  {631} (\bibinfo {year} {2020})}\BibitemShut {NoStop}%
\bibitem [{\citenamefont {Dias}\ \emph {et~al.}(2019)\citenamefont {Dias},
  \citenamefont {Noked},\ and\ \citenamefont {Silvera}}]{Dias2019}%
  \BibitemOpen
  \bibfield  {author} {\bibinfo {author} {\bibfnamefont {R.~P.}\ \bibnamefont
  {Dias}}, \bibinfo {author} {\bibfnamefont {O.}~\bibnamefont {Noked}},\ and\
  \bibinfo {author} {\bibfnamefont {I.~F.}\ \bibnamefont {Silvera}},\
  }\bibfield  {title} {\bibinfo {title} {{Quantum phase transition in solid
  hydrogen at high pressure}},\ }\href
  {https://doi.org/10.1103/PhysRevB.100.184112} {\bibfield  {journal} {\bibinfo
   {journal} {Physical Review B}\ }\textbf {\bibinfo {volume} {100}},\ \bibinfo
  {pages} {184112} (\bibinfo {year} {2019})}\BibitemShut {NoStop}%
\bibitem [{\citenamefont {Eremets}\ \emph {et~al.}(2019)\citenamefont
  {Eremets}, \citenamefont {Drozdov}, \citenamefont {Kong},\ and\ \citenamefont
  {Wang}}]{Eremets2019}%
  \BibitemOpen
  \bibfield  {author} {\bibinfo {author} {\bibfnamefont {M.~I.}\ \bibnamefont
  {Eremets}}, \bibinfo {author} {\bibfnamefont {A.~P.}\ \bibnamefont
  {Drozdov}}, \bibinfo {author} {\bibfnamefont {P.~P.}\ \bibnamefont {Kong}},\
  and\ \bibinfo {author} {\bibfnamefont {H.}~\bibnamefont {Wang}},\ }\bibfield
  {title} {\bibinfo {title} {{Semimetallic molecular hydrogen at pressure above
  350 GPa}},\ }\bibfield  {journal} {\bibinfo  {journal} {Nat. Phys.}\ }\href
  {https://doi.org/10.1038/s41567-019-0646-x} {10.1038/s41567-019-0646-x}
  (\bibinfo {year} {2019})\BibitemShut {NoStop}%
\bibitem [{\citenamefont {Rueff}\ and\ \citenamefont
  {Shukla}(2010)}]{Rueff2010}%
  \BibitemOpen
  \bibfield  {author} {\bibinfo {author} {\bibfnamefont {J.~P.}\ \bibnamefont
  {Rueff}}\ and\ \bibinfo {author} {\bibfnamefont {A.}~\bibnamefont {Shukla}},\
  }\bibfield  {title} {\bibinfo {title} {Inelastic x-ray scattering by
  electronic excitations under high pressure},\ }\href
  {https://doi.org/10.1103/RevModPhys.82.847} {\bibfield  {journal} {\bibinfo
  {journal} {Reviews of Modern Physics}\ }\textbf {\bibinfo {volume} {82}},\
  \bibinfo {pages} {847} (\bibinfo {year} {2010})}\BibitemShut {NoStop}%
\bibitem [{\citenamefont {Shen}\ and\ \citenamefont {Mao}(2017)}]{Shen2017}%
  \BibitemOpen
  \bibfield  {author} {\bibinfo {author} {\bibfnamefont {G.}~\bibnamefont
  {Shen}}\ and\ \bibinfo {author} {\bibfnamefont {H.~K.}\ \bibnamefont {Mao}},\
  }\bibfield  {title} {\bibinfo {title} {High-pressure studies with x-rays
  using diamond anvil cells},\ }\bibfield  {journal} {\bibinfo  {journal}
  {Reports on Progress in Physics}\ }\textbf {\bibinfo {volume} {80}},\ \href
  {https://doi.org/10.1088/1361-6633/80/1/016101}
  {10.1088/1361-6633/80/1/016101} (\bibinfo {year} {2017})\BibitemShut
  {NoStop}%
\bibitem [{\citenamefont {Mao}\ \emph {et~al.}(2018)\citenamefont {Mao},
  \citenamefont {Chen}, \citenamefont {Ding}, \citenamefont {Li},\ and\
  \citenamefont {Wang}}]{Mao2018}%
  \BibitemOpen
  \bibfield  {author} {\bibinfo {author} {\bibfnamefont {H.-K.}\ \bibnamefont
  {Mao}}, \bibinfo {author} {\bibfnamefont {X.-J.}\ \bibnamefont {Chen}},
  \bibinfo {author} {\bibfnamefont {Y.}~\bibnamefont {Ding}}, \bibinfo {author}
  {\bibfnamefont {B.}~\bibnamefont {Li}},\ and\ \bibinfo {author}
  {\bibfnamefont {L.}~\bibnamefont {Wang}},\ }\bibfield  {title} {\bibinfo
  {title} {Solids, liquids, and gases under high pressure},\ }\href
  {https://doi.org/10.1103/RevModPhys.90.015007} {\bibfield  {journal}
  {\bibinfo  {journal} {Reviews of Modern Physics}\ }\textbf {\bibinfo {volume}
  {90}},\ \bibinfo {pages} {015007} (\bibinfo {year} {2018})}\BibitemShut
  {NoStop}%
\bibitem [{\citenamefont {Li}\ \emph {et~al.}(2021)\citenamefont {Li},
  \citenamefont {Ding}, \citenamefont {Kim}, \citenamefont {Wang},
  \citenamefont {Weng}, \citenamefont {Yang}, \citenamefont {Yu}, \citenamefont
  {Ji}, \citenamefont {Wang}, \citenamefont {Shu}, \citenamefont {Chen},
  \citenamefont {Yang}, \citenamefont {Xiao}, \citenamefont {Chow},
  \citenamefont {Shen}, \citenamefont {Mao},\ and\ \citenamefont
  {Mao}}]{Li2021}%
  \BibitemOpen
  \bibfield  {author} {\bibinfo {author} {\bibfnamefont {B.}~\bibnamefont
  {Li}}, \bibinfo {author} {\bibfnamefont {Y.}~\bibnamefont {Ding}}, \bibinfo
  {author} {\bibfnamefont {D.~Y.}\ \bibnamefont {Kim}}, \bibinfo {author}
  {\bibfnamefont {L.}~\bibnamefont {Wang}}, \bibinfo {author} {\bibfnamefont
  {T.-C.}\ \bibnamefont {Weng}}, \bibinfo {author} {\bibfnamefont
  {W.}~\bibnamefont {Yang}}, \bibinfo {author} {\bibfnamefont {Z.}~\bibnamefont
  {Yu}}, \bibinfo {author} {\bibfnamefont {C.}~\bibnamefont {Ji}}, \bibinfo
  {author} {\bibfnamefont {J.}~\bibnamefont {Wang}}, \bibinfo {author}
  {\bibfnamefont {J.}~\bibnamefont {Shu}}, \bibinfo {author} {\bibfnamefont
  {J.}~\bibnamefont {Chen}}, \bibinfo {author} {\bibfnamefont {K.}~\bibnamefont
  {Yang}}, \bibinfo {author} {\bibfnamefont {Y.}~\bibnamefont {Xiao}}, \bibinfo
  {author} {\bibfnamefont {P.}~\bibnamefont {Chow}}, \bibinfo {author}
  {\bibfnamefont {G.}~\bibnamefont {Shen}}, \bibinfo {author} {\bibfnamefont
  {W.~L.}\ \bibnamefont {Mao}},\ and\ \bibinfo {author} {\bibfnamefont {H.-K.}\
  \bibnamefont {Mao}},\ }\bibfield  {title} {\bibinfo {title} {{Probing the
  Electronic Band Gap of Solid Hydrogen by Inelastic X-Ray Scattering up to 90
  GPa}},\ }\href {https://doi.org/10.1103/PhysRevLett.126.036402} {\bibfield
  {journal} {\bibinfo  {journal} {Physical Review Letters}\ }\textbf {\bibinfo
  {volume} {126}},\ \bibinfo {pages} {36402} (\bibinfo {year}
  {2021})}\BibitemShut {NoStop}%
\bibitem [{\citenamefont {Martin}\ \emph {et~al.}(2016)\citenamefont {Martin},
  \citenamefont {Reining},\ and\ \citenamefont {Ceperley}}]{Martin2016}%
  \BibitemOpen
  \bibfield  {author} {\bibinfo {author} {\bibfnamefont {R.~M.}\ \bibnamefont
  {Martin}}, \bibinfo {author} {\bibfnamefont {L.}~\bibnamefont {Reining}},\
  and\ \bibinfo {author} {\bibfnamefont {D.~M.}\ \bibnamefont {Ceperley}},\
  }\href {https://doi.org/10.1017/CBO9781139050807} {\emph {\bibinfo {title}
  {{Interacting Electrons}}}}\ (\bibinfo  {publisher} {Cambridge University
  Press},\ \bibinfo {address} {Cambridge},\ \bibinfo {year} {2016})\BibitemShut
  {NoStop}%
\bibitem [{\citenamefont {Dvorak}\ \emph {et~al.}(2014)\citenamefont {Dvorak},
  \citenamefont {Chen},\ and\ \citenamefont {Wu}}]{Dvorak2014}%
  \BibitemOpen
  \bibfield  {author} {\bibinfo {author} {\bibfnamefont {M.}~\bibnamefont
  {Dvorak}}, \bibinfo {author} {\bibfnamefont {X.-J.}\ \bibnamefont {Chen}},\
  and\ \bibinfo {author} {\bibfnamefont {Z.}~\bibnamefont {Wu}},\ }\bibfield
  {title} {\bibinfo {title} {{Quasiparticle energies and excitonic effects in
  dense solid hydrogen near metallization}},\ }\href
  {https://doi.org/10.1103/PhysRevB.90.035103} {\bibfield  {journal} {\bibinfo
  {journal} {Physical Review B}\ }\textbf {\bibinfo {volume} {90}},\ \bibinfo
  {pages} {035103} (\bibinfo {year} {2014})}\BibitemShut {NoStop}%
\bibitem [{\citenamefont {Lebegue}\ \emph {et~al.}(2012)\citenamefont
  {Lebegue}, \citenamefont {Araujo}, \citenamefont {Kim}, \citenamefont
  {Ramzan}, \citenamefont {Mao},\ and\ \citenamefont {Ahuja}}]{Lebegue2012}%
  \BibitemOpen
  \bibfield  {author} {\bibinfo {author} {\bibfnamefont {S.}~\bibnamefont
  {Lebegue}}, \bibinfo {author} {\bibfnamefont {C.~M.}\ \bibnamefont {Araujo}},
  \bibinfo {author} {\bibfnamefont {D.~Y.}\ \bibnamefont {Kim}}, \bibinfo
  {author} {\bibfnamefont {M.}~\bibnamefont {Ramzan}}, \bibinfo {author}
  {\bibfnamefont {H.-k.}\ \bibnamefont {Mao}},\ and\ \bibinfo {author}
  {\bibfnamefont {R.}~\bibnamefont {Ahuja}},\ }\bibfield  {title} {\bibinfo
  {title} {{Semimetallic dense hydrogen above 260 GPa}},\ }\href
  {https://doi.org/10.1073/pnas.1207065109} {\bibfield  {journal} {\bibinfo
  {journal} {Proceedings of the National Academy of Sciences}\ }\textbf
  {\bibinfo {volume} {109}},\ \bibinfo {pages} {9766} (\bibinfo {year}
  {2012})}\BibitemShut {NoStop}%
\bibitem [{\citenamefont {Azadi}\ \emph {et~al.}(2022)\citenamefont {Azadi},
  \citenamefont {Davydov},\ and\ \citenamefont {Kozik}}]{Azadi2022}%
  \BibitemOpen
  \bibfield  {author} {\bibinfo {author} {\bibfnamefont {S.}~\bibnamefont
  {Azadi}}, \bibinfo {author} {\bibfnamefont {A.}~\bibnamefont {Davydov}},\
  and\ \bibinfo {author} {\bibfnamefont {E.}~\bibnamefont {Kozik}},\ }\bibfield
   {title} {\bibinfo {title} {{$GW$ space-time method: Energy band gap of solid
  hydrogen}},\ }\href {https://doi.org/10.1103/PhysRevB.105.155136} {\bibfield
  {journal} {\bibinfo  {journal} {Physical Review B}\ }\textbf {\bibinfo
  {volume} {105}},\ \bibinfo {pages} {155136} (\bibinfo {year}
  {2022})}\BibitemShut {NoStop}%
\bibitem [{\citenamefont {Kioupakis}\ \emph {et~al.}(2008)\citenamefont
  {Kioupakis}, \citenamefont {Zhang}, \citenamefont {Cohen},\ and\
  \citenamefont {Louie}}]{Kioupakis2008}%
  \BibitemOpen
  \bibfield  {author} {\bibinfo {author} {\bibfnamefont {E.}~\bibnamefont
  {Kioupakis}}, \bibinfo {author} {\bibfnamefont {P.}~\bibnamefont {Zhang}},
  \bibinfo {author} {\bibfnamefont {M.~L.}\ \bibnamefont {Cohen}},\ and\
  \bibinfo {author} {\bibfnamefont {S.~G.}\ \bibnamefont {Louie}},\ }\bibfield
  {title} {\bibinfo {title} {{GW quasiparticle corrections to the LDA+U GGA+U
  electronic structure of bcc hydrogen}},\ }\href
  {https://doi.org/10.1103/PhysRevB.77.155114} {\bibfield  {journal} {\bibinfo
  {journal} {Physical Review B - Condensed Matter and Materials Physics}\
  }\textbf {\bibinfo {volume} {77}},\ \bibinfo {pages} {1} (\bibinfo {year}
  {2008})}\BibitemShut {NoStop}%
\bibitem [{\citenamefont {Gorelov}\ \emph
  {et~al.}(2020{\natexlab{a}})\citenamefont {Gorelov}, \citenamefont
  {Holzmann}, \citenamefont {Ceperley},\ and\ \citenamefont
  {Pierleoni}}]{Gorelov2020}%
  \BibitemOpen
  \bibfield  {author} {\bibinfo {author} {\bibfnamefont {V.}~\bibnamefont
  {Gorelov}}, \bibinfo {author} {\bibfnamefont {M.}~\bibnamefont {Holzmann}},
  \bibinfo {author} {\bibfnamefont {D.~M.}\ \bibnamefont {Ceperley}},\ and\
  \bibinfo {author} {\bibfnamefont {C.}~\bibnamefont {Pierleoni}},\ }\bibfield
  {title} {\bibinfo {title} {{Energy Gap Closure of Crystalline Molecular
  Hydrogen with Pressure}},\ }\href
  {https://doi.org/10.1103/PhysRevLett.124.116401} {\bibfield  {journal}
  {\bibinfo  {journal} {Physical Review Letters}\ }\textbf {\bibinfo {volume}
  {124}},\ \bibinfo {pages} {116401} (\bibinfo {year} {2020}{\natexlab{a}})},\
  \Eprint {https://arxiv.org/abs/1911.06135} {arXiv:1911.06135} \BibitemShut
  {NoStop}%
\bibitem [{\citenamefont {Gorelov}\ \emph
  {et~al.}(2020{\natexlab{b}})\citenamefont {Gorelov}, \citenamefont
  {Ceperley}, \citenamefont {Holzmann},\ and\ \citenamefont
  {Pierleoni}}]{Gorelov2020a}%
  \BibitemOpen
  \bibfield  {author} {\bibinfo {author} {\bibfnamefont {V.}~\bibnamefont
  {Gorelov}}, \bibinfo {author} {\bibfnamefont {D.~M.}\ \bibnamefont
  {Ceperley}}, \bibinfo {author} {\bibfnamefont {M.}~\bibnamefont {Holzmann}},\
  and\ \bibinfo {author} {\bibfnamefont {C.}~\bibnamefont {Pierleoni}},\
  }\bibfield  {title} {\bibinfo {title} {{Electronic energy gap closure and
  metal-insulator transition in dense liquid hydrogen}},\ }\href
  {https://doi.org/10.1103/PhysRevB.102.195133} {\bibfield  {journal} {\bibinfo
   {journal} {Physical Review B}\ }\textbf {\bibinfo {volume} {102}},\ \bibinfo
  {pages} {195133} (\bibinfo {year} {2020}{\natexlab{b}})},\ \Eprint
  {https://arxiv.org/abs/2009.00652} {arXiv:2009.00652} \BibitemShut {NoStop}%
\bibitem [{\citenamefont {Azadi}\ \emph {et~al.}(2017)\citenamefont {Azadi},
  \citenamefont {Drummond},\ and\ \citenamefont {Foulkes}}]{Azadi2017}%
  \BibitemOpen
  \bibfield  {author} {\bibinfo {author} {\bibfnamefont {S.}~\bibnamefont
  {Azadi}}, \bibinfo {author} {\bibfnamefont {N.~D.}\ \bibnamefont
  {Drummond}},\ and\ \bibinfo {author} {\bibfnamefont {W.~M.~C.}\ \bibnamefont
  {Foulkes}},\ }\bibfield  {title} {\bibinfo {title} {{Nature of the
  metallization transition in solid hydrogen}},\ }\href@noop {} {\bibfield
  {journal} {\bibinfo  {journal} {Phys. Rev. B}\ }\textbf {\bibinfo {volume}
  {95}},\ \bibinfo {pages} {035142} (\bibinfo {year} {2017})}\BibitemShut
  {NoStop}%
\bibitem [{\citenamefont {Monacelli}\ \emph {et~al.}(2023)\citenamefont
  {Monacelli}, \citenamefont {Casula}, \citenamefont {Nakano}, \citenamefont
  {Sorella},\ and\ \citenamefont {Mauri}}]{Monacelli2023}%
  \BibitemOpen
  \bibfield  {author} {\bibinfo {author} {\bibfnamefont {L.}~\bibnamefont
  {Monacelli}}, \bibinfo {author} {\bibfnamefont {M.}~\bibnamefont {Casula}},
  \bibinfo {author} {\bibfnamefont {K.}~\bibnamefont {Nakano}}, \bibinfo
  {author} {\bibfnamefont {S.}~\bibnamefont {Sorella}},\ and\ \bibinfo {author}
  {\bibfnamefont {F.}~\bibnamefont {Mauri}},\ }\bibfield  {title} {\bibinfo
  {title} {{Quantum phase diagram of high-pressure hydrogen}},\ }\bibfield
  {journal} {\bibinfo  {journal} {Nature Physics}\ }\href
  {https://doi.org/10.1038/s41567-023-01960-5} {10.1038/s41567-023-01960-5}
  (\bibinfo {year} {2023}),\ \Eprint {https://arxiv.org/abs/2202.05740}
  {arXiv:2202.05740} \BibitemShut {NoStop}%
\bibitem [{\citenamefont {Niu}\ \emph {et~al.}(2023)\citenamefont {Niu},
  \citenamefont {Yang}, \citenamefont {Jensen}, \citenamefont {Holzmann},
  \citenamefont {Pierleoni},\ and\ \citenamefont {Ceperley}}]{Niu2023}%
  \BibitemOpen
  \bibfield  {author} {\bibinfo {author} {\bibfnamefont {H.}~\bibnamefont
  {Niu}}, \bibinfo {author} {\bibfnamefont {Y.}~\bibnamefont {Yang}}, \bibinfo
  {author} {\bibfnamefont {S.}~\bibnamefont {Jensen}}, \bibinfo {author}
  {\bibfnamefont {M.}~\bibnamefont {Holzmann}}, \bibinfo {author}
  {\bibfnamefont {C.}~\bibnamefont {Pierleoni}},\ and\ \bibinfo {author}
  {\bibfnamefont {D.~M.}\ \bibnamefont {Ceperley}},\ }\bibfield  {title}
  {\bibinfo {title} {Stable solid molecular hydrogen above 900k from a
  machine-learned potential trained with diffusion quantum monte carlo},\
  }\href {https://doi.org/10.1103/PhysRevLett.130.076102} {\bibfield  {journal}
  {\bibinfo  {journal} {Physical Review Letters}\ }\textbf {\bibinfo {volume}
  {130}},\ \bibinfo {pages} {76102} (\bibinfo {year} {2023})}\BibitemShut
  {NoStop}%
\bibitem [{\citenamefont {Clay}\ \emph {et~al.}(2014)\citenamefont {Clay},
  \citenamefont {McMinis}, \citenamefont {McMahon}, \citenamefont {Pierleoni},
  \citenamefont {Ceperley},\ and\ \citenamefont {Morales}}]{Clay2014}%
  \BibitemOpen
  \bibfield  {author} {\bibinfo {author} {\bibfnamefont {R.~C.}\ \bibnamefont
  {Clay}}, \bibinfo {author} {\bibfnamefont {J.}~\bibnamefont {McMinis}},
  \bibinfo {author} {\bibfnamefont {J.~M.}\ \bibnamefont {McMahon}}, \bibinfo
  {author} {\bibfnamefont {C.}~\bibnamefont {Pierleoni}}, \bibinfo {author}
  {\bibfnamefont {D.~M.}\ \bibnamefont {Ceperley}},\ and\ \bibinfo {author}
  {\bibfnamefont {M.~A.}\ \bibnamefont {Morales}},\ }\bibfield  {title}
  {\bibinfo {title} {{Benchmarking exchange-correlation functionals for
  hydrogen at high pressures using quantum Monte Carlo}},\ }\href
  {https://doi.org/10.1103/PhysRevB.89.184106} {\bibfield  {journal} {\bibinfo
  {journal} {Phys. Rev. B}\ }\textbf {\bibinfo {volume} {89}},\ \bibinfo
  {pages} {184106} (\bibinfo {year} {2014})}\BibitemShut {NoStop}%
\bibitem [{\citenamefont {Gorelov}\ \emph {et~al.}(2019)\citenamefont
  {Gorelov}, \citenamefont {Pierleoni},\ and\ \citenamefont
  {Ceperley}}]{Gorelov2019}%
  \BibitemOpen
  \bibfield  {author} {\bibinfo {author} {\bibfnamefont {V.}~\bibnamefont
  {Gorelov}}, \bibinfo {author} {\bibfnamefont {C.}~\bibnamefont {Pierleoni}},\
  and\ \bibinfo {author} {\bibfnamefont {D.~M.}\ \bibnamefont {Ceperley}},\
  }\bibfield  {title} {\bibinfo {title} {{Benchmarking vdW-DF first-principles
  predictions against Coupled Electron–Ion Monte Carlo for high-pressure
  liquid hydrogen}},\ }\href {https://doi.org/10.1002/ctpp.201800185}
  {\bibfield  {journal} {\bibinfo  {journal} {Contributions to Plasma Physics}\
  }\textbf {\bibinfo {volume} {59}},\ \bibinfo {pages} {1} (\bibinfo {year}
  {2019})},\ \Eprint {https://arxiv.org/abs/1812.07818} {arXiv:1812.07818}
  \BibitemShut {NoStop}%
\bibitem [{SM()}]{SM}%
  \BibitemOpen
  \href@noop {} {\bibinfo {title} {{See Supplemental Material at [URL will be
  inserted by publisher] for the details of the calculations performed,
  theoretical background of the methods used, and convergence
  studies.}}}\BibitemShut {Stop}%
\bibitem [{\citenamefont {Holzmann}\ \emph {et~al.}(2003)\citenamefont
  {Holzmann}, \citenamefont {Ceperley}, \citenamefont {Pierleoni},\ and\
  \citenamefont {Esler}}]{Holzmann2003}%
  \BibitemOpen
  \bibfield  {author} {\bibinfo {author} {\bibfnamefont {M.}~\bibnamefont
  {Holzmann}}, \bibinfo {author} {\bibfnamefont {D.~M.}\ \bibnamefont
  {Ceperley}}, \bibinfo {author} {\bibfnamefont {C.}~\bibnamefont
  {Pierleoni}},\ and\ \bibinfo {author} {\bibfnamefont {K.}~\bibnamefont
  {Esler}},\ }\bibfield  {title} {\bibinfo {title} {{Backflow correlations for
  the electron gas and metallic hydrogen}},\ }\href
  {https://doi.org/10.1103/PhysRevE.68.046707} {\bibfield  {journal} {\bibinfo
  {journal} {Physical Review E}\ }\textbf {\bibinfo {volume} {68}},\ \bibinfo
  {pages} {046707} (\bibinfo {year} {2003})},\ \Eprint
  {https://arxiv.org/abs/0304165} {arXiv:0304165 [cond-mat]} \BibitemShut
  {NoStop}%
\bibitem [{\citenamefont {Pierleoni}\ \emph {et~al.}(2008)\citenamefont
  {Pierleoni}, \citenamefont {Delaney}, \citenamefont {Morales}, \citenamefont
  {Ceperley},\ and\ \citenamefont {Holzmann}}]{Pierleoni2008}%
  \BibitemOpen
  \bibfield  {author} {\bibinfo {author} {\bibfnamefont {C.}~\bibnamefont
  {Pierleoni}}, \bibinfo {author} {\bibfnamefont {K.~T.}\ \bibnamefont
  {Delaney}}, \bibinfo {author} {\bibfnamefont {M.~A.}\ \bibnamefont
  {Morales}}, \bibinfo {author} {\bibfnamefont {D.~M.}\ \bibnamefont
  {Ceperley}},\ and\ \bibinfo {author} {\bibfnamefont {M.}~\bibnamefont
  {Holzmann}},\ }\bibfield  {title} {\bibinfo {title} {{Trial wave functions
  for high-pressure metallic hydrogen}},\ }\href
  {https://doi.org/10.1016/j.cpc.2008.01.041} {\bibfield  {journal} {\bibinfo
  {journal} {Computer Physics Communications}\ }\textbf {\bibinfo {volume}
  {179}},\ \bibinfo {pages} {89} (\bibinfo {year} {2008})},\ \Eprint
  {https://arxiv.org/abs/0712.0161} {arXiv:0712.0161} \BibitemShut {NoStop}%
\bibitem [{\citenamefont {Giannozzi}\ \emph {et~al.}(2009)\citenamefont
  {Giannozzi}, \citenamefont {Baroni}, \citenamefont {Bonini}, \citenamefont
  {Calandra}, \citenamefont {Car}, \citenamefont {Cavazzoni}, \citenamefont
  {Ceresoli}, \citenamefont {Chiarotti}, \citenamefont {Cococcioni},
  \citenamefont {Dabo}, \citenamefont {Corso}, \citenamefont {de~Gironcoli},
  \citenamefont {Fabris}, \citenamefont {Fratesi}, \citenamefont {Gebauer},
  \citenamefont {Gerstmann}, \citenamefont {Gougoussis}, \citenamefont
  {Kokalj}, \citenamefont {Lazzeri}, \citenamefont {Martin-Samos},
  \citenamefont {Marzari}, \citenamefont {Mauri}, \citenamefont {Mazzarello},
  \citenamefont {Paolini}, \citenamefont {Pasquarello}, \citenamefont
  {Paulatto}, \citenamefont {Sbraccia}, \citenamefont {Scandolo}, \citenamefont
  {Sclauzero}, \citenamefont {Seitsonen}, \citenamefont {Smogunov},
  \citenamefont {Umari},\ and\ \citenamefont {Wentzcovitch}}]{QE2009}%
  \BibitemOpen
  \bibfield  {author} {\bibinfo {author} {\bibfnamefont {P.}~\bibnamefont
  {Giannozzi}}, \bibinfo {author} {\bibfnamefont {S.}~\bibnamefont {Baroni}},
  \bibinfo {author} {\bibfnamefont {N.}~\bibnamefont {Bonini}}, \bibinfo
  {author} {\bibfnamefont {M.}~\bibnamefont {Calandra}}, \bibinfo {author}
  {\bibfnamefont {R.}~\bibnamefont {Car}}, \bibinfo {author} {\bibfnamefont
  {C.}~\bibnamefont {Cavazzoni}}, \bibinfo {author} {\bibfnamefont
  {D.}~\bibnamefont {Ceresoli}}, \bibinfo {author} {\bibfnamefont {G.~L.}\
  \bibnamefont {Chiarotti}}, \bibinfo {author} {\bibfnamefont {M.}~\bibnamefont
  {Cococcioni}}, \bibinfo {author} {\bibfnamefont {I.}~\bibnamefont {Dabo}},
  \bibinfo {author} {\bibfnamefont {A.~D.}\ \bibnamefont {Corso}}, \bibinfo
  {author} {\bibfnamefont {S.}~\bibnamefont {de~Gironcoli}}, \bibinfo {author}
  {\bibfnamefont {S.}~\bibnamefont {Fabris}}, \bibinfo {author} {\bibfnamefont
  {G.}~\bibnamefont {Fratesi}}, \bibinfo {author} {\bibfnamefont
  {R.}~\bibnamefont {Gebauer}}, \bibinfo {author} {\bibfnamefont
  {U.}~\bibnamefont {Gerstmann}}, \bibinfo {author} {\bibfnamefont
  {C.}~\bibnamefont {Gougoussis}}, \bibinfo {author} {\bibfnamefont
  {A.}~\bibnamefont {Kokalj}}, \bibinfo {author} {\bibfnamefont
  {M.}~\bibnamefont {Lazzeri}}, \bibinfo {author} {\bibfnamefont
  {L.}~\bibnamefont {Martin-Samos}}, \bibinfo {author} {\bibfnamefont
  {N.}~\bibnamefont {Marzari}}, \bibinfo {author} {\bibfnamefont
  {F.}~\bibnamefont {Mauri}}, \bibinfo {author} {\bibfnamefont
  {R.}~\bibnamefont {Mazzarello}}, \bibinfo {author} {\bibfnamefont
  {S.}~\bibnamefont {Paolini}}, \bibinfo {author} {\bibfnamefont
  {A.}~\bibnamefont {Pasquarello}}, \bibinfo {author} {\bibfnamefont
  {L.}~\bibnamefont {Paulatto}}, \bibinfo {author} {\bibfnamefont
  {C.}~\bibnamefont {Sbraccia}}, \bibinfo {author} {\bibfnamefont
  {S.}~\bibnamefont {Scandolo}}, \bibinfo {author} {\bibfnamefont
  {G.}~\bibnamefont {Sclauzero}}, \bibinfo {author} {\bibfnamefont {A.~P.}\
  \bibnamefont {Seitsonen}}, \bibinfo {author} {\bibfnamefont {A.}~\bibnamefont
  {Smogunov}}, \bibinfo {author} {\bibfnamefont {P.}~\bibnamefont {Umari}},\
  and\ \bibinfo {author} {\bibfnamefont {R.~M.}\ \bibnamefont {Wentzcovitch}},\
  }\bibfield  {title} {\bibinfo {title} {{QUANTUM} {ESPRESSO}: a modular and
  open-source software project for quantum simulations of materials},\ }\href
  {https://doi.org/10.1088/0953-8984/21/39/395502} {\bibfield  {journal}
  {\bibinfo  {journal} {Journal of Physics: Condensed Matter}\ }\textbf
  {\bibinfo {volume} {21}},\ \bibinfo {pages} {395502} (\bibinfo {year}
  {2009})}\BibitemShut {NoStop}%
\bibitem [{\citenamefont {Giannozzi}\ \emph {et~al.}(2017)\citenamefont
  {Giannozzi}, \citenamefont {Andreussi}, \citenamefont {Brumme}, \citenamefont
  {Bunau}, \citenamefont {Nardelli}, \citenamefont {Calandra}, \citenamefont
  {Car}, \citenamefont {Cavazzoni}, \citenamefont {Ceresoli}, \citenamefont
  {Cococcioni}, \citenamefont {Colonna}, \citenamefont {Carnimeo},
  \citenamefont {Corso}, \citenamefont {de~Gironcoli}, \citenamefont {Delugas},
  \citenamefont {DiStasio}, \citenamefont {Ferretti}, \citenamefont {Floris},
  \citenamefont {Fratesi}, \citenamefont {Fugallo}, \citenamefont {Gebauer},
  \citenamefont {Gerstmann}, \citenamefont {Giustino}, \citenamefont {Gorni},
  \citenamefont {Jia}, \citenamefont {Kawamura}, \citenamefont {Ko},
  \citenamefont {Kokalj}, \citenamefont {KÃŒ{\c{c}}ÃŒkbenli}, \citenamefont
  {Lazzeri}, \citenamefont {Marsili}, \citenamefont {Marzari}, \citenamefont
  {Mauri}, \citenamefont {Nguyen}, \citenamefont {Nguyen}, \citenamefont {de-la
  Roza}, \citenamefont {Paulatto}, \citenamefont {Ponc{\'{e}}}, \citenamefont
  {Rocca}, \citenamefont {Sabatini}, \citenamefont {Santra}, \citenamefont
  {Schlipf}, \citenamefont {Seitsonen}, \citenamefont {Smogunov}, \citenamefont
  {Timrov}, \citenamefont {Thonhauser}, \citenamefont {Umari}, \citenamefont
  {Vast}, \citenamefont {Wu},\ and\ \citenamefont {Baroni}}]{QE2017}%
  \BibitemOpen
  \bibfield  {author} {\bibinfo {author} {\bibfnamefont {P.}~\bibnamefont
  {Giannozzi}}, \bibinfo {author} {\bibfnamefont {O.}~\bibnamefont
  {Andreussi}}, \bibinfo {author} {\bibfnamefont {T.}~\bibnamefont {Brumme}},
  \bibinfo {author} {\bibfnamefont {O.}~\bibnamefont {Bunau}}, \bibinfo
  {author} {\bibfnamefont {M.~B.}\ \bibnamefont {Nardelli}}, \bibinfo {author}
  {\bibfnamefont {M.}~\bibnamefont {Calandra}}, \bibinfo {author}
  {\bibfnamefont {R.}~\bibnamefont {Car}}, \bibinfo {author} {\bibfnamefont
  {C.}~\bibnamefont {Cavazzoni}}, \bibinfo {author} {\bibfnamefont
  {D.}~\bibnamefont {Ceresoli}}, \bibinfo {author} {\bibfnamefont
  {M.}~\bibnamefont {Cococcioni}}, \bibinfo {author} {\bibfnamefont
  {N.}~\bibnamefont {Colonna}}, \bibinfo {author} {\bibfnamefont
  {I.}~\bibnamefont {Carnimeo}}, \bibinfo {author} {\bibfnamefont {A.~D.}\
  \bibnamefont {Corso}}, \bibinfo {author} {\bibfnamefont {S.}~\bibnamefont
  {de~Gironcoli}}, \bibinfo {author} {\bibfnamefont {P.}~\bibnamefont
  {Delugas}}, \bibinfo {author} {\bibfnamefont {R.~A.}\ \bibnamefont
  {DiStasio}}, \bibinfo {author} {\bibfnamefont {A.}~\bibnamefont {Ferretti}},
  \bibinfo {author} {\bibfnamefont {A.}~\bibnamefont {Floris}}, \bibinfo
  {author} {\bibfnamefont {G.}~\bibnamefont {Fratesi}}, \bibinfo {author}
  {\bibfnamefont {G.}~\bibnamefont {Fugallo}}, \bibinfo {author} {\bibfnamefont
  {R.}~\bibnamefont {Gebauer}}, \bibinfo {author} {\bibfnamefont
  {U.}~\bibnamefont {Gerstmann}}, \bibinfo {author} {\bibfnamefont
  {F.}~\bibnamefont {Giustino}}, \bibinfo {author} {\bibfnamefont
  {T.}~\bibnamefont {Gorni}}, \bibinfo {author} {\bibfnamefont
  {J.}~\bibnamefont {Jia}}, \bibinfo {author} {\bibfnamefont {M.}~\bibnamefont
  {Kawamura}}, \bibinfo {author} {\bibfnamefont {H.-Y.}\ \bibnamefont {Ko}},
  \bibinfo {author} {\bibfnamefont {A.}~\bibnamefont {Kokalj}}, \bibinfo
  {author} {\bibfnamefont {E.}~\bibnamefont {KÃŒ{\c{c}}ÃŒkbenli}}, \bibinfo
  {author} {\bibfnamefont {M.}~\bibnamefont {Lazzeri}}, \bibinfo {author}
  {\bibfnamefont {M.}~\bibnamefont {Marsili}}, \bibinfo {author} {\bibfnamefont
  {N.}~\bibnamefont {Marzari}}, \bibinfo {author} {\bibfnamefont
  {F.}~\bibnamefont {Mauri}}, \bibinfo {author} {\bibfnamefont {N.~L.}\
  \bibnamefont {Nguyen}}, \bibinfo {author} {\bibfnamefont {H.-V.}\
  \bibnamefont {Nguyen}}, \bibinfo {author} {\bibfnamefont {A.~O.}\
  \bibnamefont {de-la Roza}}, \bibinfo {author} {\bibfnamefont
  {L.}~\bibnamefont {Paulatto}}, \bibinfo {author} {\bibfnamefont
  {S.}~\bibnamefont {Ponc{\'{e}}}}, \bibinfo {author} {\bibfnamefont
  {D.}~\bibnamefont {Rocca}}, \bibinfo {author} {\bibfnamefont
  {R.}~\bibnamefont {Sabatini}}, \bibinfo {author} {\bibfnamefont
  {B.}~\bibnamefont {Santra}}, \bibinfo {author} {\bibfnamefont
  {M.}~\bibnamefont {Schlipf}}, \bibinfo {author} {\bibfnamefont {A.~P.}\
  \bibnamefont {Seitsonen}}, \bibinfo {author} {\bibfnamefont {A.}~\bibnamefont
  {Smogunov}}, \bibinfo {author} {\bibfnamefont {I.}~\bibnamefont {Timrov}},
  \bibinfo {author} {\bibfnamefont {T.}~\bibnamefont {Thonhauser}}, \bibinfo
  {author} {\bibfnamefont {P.}~\bibnamefont {Umari}}, \bibinfo {author}
  {\bibfnamefont {N.}~\bibnamefont {Vast}}, \bibinfo {author} {\bibfnamefont
  {X.}~\bibnamefont {Wu}},\ and\ \bibinfo {author} {\bibfnamefont
  {S.}~\bibnamefont {Baroni}},\ }\bibfield  {title} {\bibinfo {title} {Advanced
  capabilities for materials modelling with quantum {ESPRESSO}},\ }\href
  {https://doi.org/10.1088/1361-648x/aa8f79} {\bibfield  {journal} {\bibinfo
  {journal} {Journal of Physics: Condensed Matter}\ }\textbf {\bibinfo {volume}
  {29}},\ \bibinfo {pages} {465901} (\bibinfo {year} {2017})}\BibitemShut
  {NoStop}%
\bibitem [{\citenamefont {Morales}\ \emph {et~al.}(2013)\citenamefont
  {Morales}, \citenamefont {Clay}, \citenamefont {Pierleoni},\ and\
  \citenamefont {Ceperley}}]{Morales2013}%
  \BibitemOpen
  \bibfield  {author} {\bibinfo {author} {\bibfnamefont {M.}~\bibnamefont
  {Morales}}, \bibinfo {author} {\bibfnamefont {R.}~\bibnamefont {Clay}},
  \bibinfo {author} {\bibfnamefont {C.}~\bibnamefont {Pierleoni}},\ and\
  \bibinfo {author} {\bibfnamefont {D.}~\bibnamefont {Ceperley}},\ }\bibfield
  {title} {\bibinfo {title} {{First Principles Methods: A Perspective from
  Quantum Monte Carlo}},\ }\href {https://doi.org/10.3390/e16010287} {\bibfield
   {journal} {\bibinfo  {journal} {Entropy}\ }\textbf {\bibinfo {volume}
  {16}},\ \bibinfo {pages} {287} (\bibinfo {year} {2013})}\BibitemShut
  {NoStop}%
\bibitem [{\citenamefont {Pierleoni}\ \emph {et~al.}(2016)\citenamefont
  {Pierleoni}, \citenamefont {Morales}, \citenamefont {Rillo}, \citenamefont
  {Holzmann},\ and\ \citenamefont {Ceperley}}]{Pierleoni2016}%
  \BibitemOpen
  \bibfield  {author} {\bibinfo {author} {\bibfnamefont {C.}~\bibnamefont
  {Pierleoni}}, \bibinfo {author} {\bibfnamefont {M.~a.}\ \bibnamefont
  {Morales}}, \bibinfo {author} {\bibfnamefont {G.}~\bibnamefont {Rillo}},
  \bibinfo {author} {\bibfnamefont {M.}~\bibnamefont {Holzmann}},\ and\
  \bibinfo {author} {\bibfnamefont {D.~M.}\ \bibnamefont {Ceperley}},\
  }\bibfield  {title} {\bibinfo {title} {{Liquid–liquid phase transition in
  hydrogen by coupled electron–ion Monte Carlo simulations}},\ }\href
  {https://doi.org/10.1073/pnas.1603853113} {\bibfield  {journal} {\bibinfo
  {journal} {Proceedings of the National Academy of Sciences}\ }\textbf
  {\bibinfo {volume} {113}},\ \bibinfo {pages} {4954} (\bibinfo {year}
  {2016})}\BibitemShut {NoStop}%
\bibitem [{\citenamefont {Williams}(1951)}]{Williams1951}%
  \BibitemOpen
  \bibfield  {author} {\bibinfo {author} {\bibfnamefont {F.~E.}\ \bibnamefont
  {Williams}},\ }\bibfield  {title} {\bibinfo {title} {{An absolute theory of
  solid-state luminescence}},\ }\href {https://doi.org/10.1063/1.1748247}
  {\bibfield  {journal} {\bibinfo  {journal} {The Journal of Chemical Physics}\
  }\textbf {\bibinfo {volume} {19}},\ \bibinfo {pages} {457} (\bibinfo {year}
  {1951})}\BibitemShut {NoStop}%
\bibitem [{\citenamefont {Lax}(1952)}]{Lax1952}%
  \BibitemOpen
  \bibfield  {author} {\bibinfo {author} {\bibfnamefont {M.}~\bibnamefont
  {Lax}},\ }\bibfield  {title} {\bibinfo {title} {{The franck-condon principle
  and its application to crystals}},\ }\href
  {https://doi.org/10.1063/1.1700283} {\bibfield  {journal} {\bibinfo
  {journal} {The Journal of Chemical Physics}\ }\textbf {\bibinfo {volume}
  {20}},\ \bibinfo {pages} {1752} (\bibinfo {year} {1952})}\BibitemShut
  {NoStop}%
\bibitem [{\citenamefont {Constable}\ \emph {et~al.}(1975)\citenamefont
  {Constable}, \citenamefont {Clark},\ and\ \citenamefont
  {Gaines}}]{Constable1975}%
  \BibitemOpen
  \bibfield  {author} {\bibinfo {author} {\bibfnamefont {J.~H.}\ \bibnamefont
  {Constable}}, \bibinfo {author} {\bibfnamefont {C.~F.}\ \bibnamefont
  {Clark}},\ and\ \bibinfo {author} {\bibfnamefont {J.~R.}\ \bibnamefont
  {Gaines}},\ }\bibfield  {title} {\bibinfo {title} {{The dielectric constant
  of H2, D2, and HD in the condensed phases}},\ }\href
  {https://doi.org/10.1007/BF01141613} {\bibfield  {journal} {\bibinfo
  {journal} {Journal of Low Temperature Physics}\ }\textbf {\bibinfo {volume}
  {21}},\ \bibinfo {pages} {599} (\bibinfo {year} {1975})}\BibitemShut
  {NoStop}%
\bibitem [{\citenamefont {Annaberdiyev}\ \emph {et~al.}(2021)\citenamefont
  {Annaberdiyev}, \citenamefont {Wang}, \citenamefont {Melton}, \citenamefont
  {Bennett},\ and\ \citenamefont {Mitas}}]{Lubos21}%
  \BibitemOpen
  \bibfield  {author} {\bibinfo {author} {\bibfnamefont {A.}~\bibnamefont
  {Annaberdiyev}}, \bibinfo {author} {\bibfnamefont {G.}~\bibnamefont {Wang}},
  \bibinfo {author} {\bibfnamefont {C.~A.}\ \bibnamefont {Melton}}, \bibinfo
  {author} {\bibfnamefont {M.~C.}\ \bibnamefont {Bennett}},\ and\ \bibinfo
  {author} {\bibfnamefont {L.}~\bibnamefont {Mitas}},\ }\href
  {https://doi.org/10.1103/PhysRevB.103.205206} {\bibfield  {journal} {\bibinfo
   {journal} {Phys. Rev. B}\ }\textbf {\bibinfo {volume} {103}},\ \bibinfo
  {pages} {205206} (\bibinfo {year} {2021})}\BibitemShut {NoStop}%
\bibitem [{\citenamefont {Yang}\ \emph {et~al.}(2020)\citenamefont {Yang},
  \citenamefont {Gorelov}, \citenamefont {Pierleoni}, \citenamefont
  {Ceperley},\ and\ \citenamefont {Holzmann}}]{Yang2020}%
  \BibitemOpen
  \bibfield  {author} {\bibinfo {author} {\bibfnamefont {Y.}~\bibnamefont
  {Yang}}, \bibinfo {author} {\bibfnamefont {V.}~\bibnamefont {Gorelov}},
  \bibinfo {author} {\bibfnamefont {C.}~\bibnamefont {Pierleoni}}, \bibinfo
  {author} {\bibfnamefont {D.~M.}\ \bibnamefont {Ceperley}},\ and\ \bibinfo
  {author} {\bibfnamefont {M.}~\bibnamefont {Holzmann}},\ }\bibfield  {title}
  {\bibinfo {title} {{Electronic band gaps from Quantum Monte Carlo methods}},\
  }\href {https://doi.org/10.1103/PhysRevB.101.085115} {\bibfield  {journal}
  {\bibinfo  {journal} {Physical Review B}\ }\textbf {\bibinfo {volume}
  {101}},\ \bibinfo {pages} {85115} (\bibinfo {year} {2020})},\ \Eprint
  {https://arxiv.org/abs/1910.07531} {arXiv:1910.07531} \BibitemShut {NoStop}%
\bibitem [{\citenamefont {Ceperley}\ and\ \citenamefont
  {Alder}(1987)}]{Ceperley1987}%
  \BibitemOpen
  \bibfield  {author} {\bibinfo {author} {\bibfnamefont {D.~M.}\ \bibnamefont
  {Ceperley}}\ and\ \bibinfo {author} {\bibfnamefont {B.~J.}\ \bibnamefont
  {Alder}},\ }\bibfield  {title} {\bibinfo {title} {{Ground state of solid
  hydrogen at high pressures}},\ }\href
  {https://doi.org/10.1103/PhysRevB.36.2092} {\bibfield  {journal} {\bibinfo
  {journal} {Physical Review B}\ }\textbf {\bibinfo {volume} {36}},\ \bibinfo
  {pages} {2092} (\bibinfo {year} {1987})}\BibitemShut {NoStop}%
\bibitem [{\citenamefont {Onida}\ \emph {et~al.}(2002)\citenamefont {Onida},
  \citenamefont {Reining},\ and\ \citenamefont {Rubio}}]{Onida2002}%
  \BibitemOpen
  \bibfield  {author} {\bibinfo {author} {\bibfnamefont {G.}~\bibnamefont
  {Onida}}, \bibinfo {author} {\bibfnamefont {L.}~\bibnamefont {Reining}},\
  and\ \bibinfo {author} {\bibfnamefont {A.}~\bibnamefont {Rubio}},\ }\bibfield
   {title} {\bibinfo {title} {{Electronic excitations: density-functional
  versus many-body Green's-function approaches}},\ }\href
  {https://doi.org/10.1103/RevModPhys.74.601} {\bibfield  {journal} {\bibinfo
  {journal} {Reviews of Modern Physics}\ }\textbf {\bibinfo {volume} {74}},\
  \bibinfo {pages} {601} (\bibinfo {year} {2002})}\BibitemShut {NoStop}%
\bibitem [{\citenamefont {Strinati}(1988)}]{Strinati1988}%
  \BibitemOpen
  \bibfield  {author} {\bibinfo {author} {\bibfnamefont {G.}~\bibnamefont
  {Strinati}},\ }\href {http://dx.doi.org/10.1007/BF0272596} {\bibfield
  {journal} {\bibinfo  {journal} {Rivista del Nuovo Cimento}\ }\textbf
  {\bibinfo {volume} {11}},\ \bibinfo {pages} {1} (\bibinfo {year} {1988})},\
  \bibinfo {note} {and references therein}\BibitemShut {NoStop}%
\bibitem [{\citenamefont {Hedin}(1965)}]{Hedin1965}%
  \BibitemOpen
  \bibfield  {author} {\bibinfo {author} {\bibfnamefont {L.}~\bibnamefont
  {Hedin}},\ }\bibfield  {title} {\bibinfo {title} {New method for calculating
  the one-particle green's function with application to the electron-gas
  problem},\ }\href {https://doi.org/10.1103/PhysRev.139.A796} {\bibfield
  {journal} {\bibinfo  {journal} {Phys. Rev.}\ }\textbf {\bibinfo {volume}
  {139}},\ \bibinfo {pages} {A796} (\bibinfo {year} {1965})}\BibitemShut
  {NoStop}%
\bibitem [{\citenamefont {Adler}(1962)}]{Adler1962}%
  \BibitemOpen
  \bibfield  {author} {\bibinfo {author} {\bibfnamefont {S.~L.}\ \bibnamefont
  {Adler}},\ }\bibfield  {title} {\bibinfo {title} {Quantum theory of the
  dielectric constant in real solids},\ }\href
  {https://doi.org/10.1103/PhysRev.126.413} {\bibfield  {journal} {\bibinfo
  {journal} {Phys. Rev.}\ }\textbf {\bibinfo {volume} {126}},\ \bibinfo {pages}
  {413} (\bibinfo {year} {1962})}\BibitemShut {NoStop}%
\bibitem [{\citenamefont {Wiser}(1963)}]{Wiser1963}%
  \BibitemOpen
  \bibfield  {author} {\bibinfo {author} {\bibfnamefont {N.}~\bibnamefont
  {Wiser}},\ }\bibfield  {title} {\bibinfo {title} {Dielectric constant with
  local field effects included},\ }\href
  {https://doi.org/10.1103/PhysRev.129.62} {\bibfield  {journal} {\bibinfo
  {journal} {Phys. Rev.}\ }\textbf {\bibinfo {volume} {129}},\ \bibinfo {pages}
  {62} (\bibinfo {year} {1963})}\BibitemShut {NoStop}%
\bibitem [{\citenamefont {Hybertsen}\ and\ \citenamefont
  {Louie}(1986)}]{Hybertsen1986}%
  \BibitemOpen
  \bibfield  {author} {\bibinfo {author} {\bibfnamefont {M.~S.}\ \bibnamefont
  {Hybertsen}}\ and\ \bibinfo {author} {\bibfnamefont {S.~G.}\ \bibnamefont
  {Louie}},\ }\bibfield  {title} {\bibinfo {title} {Electron correlation in
  semiconductors and insulators: Band gaps and quasiparticle energies},\ }\href
  {https://doi.org/10.1103/PhysRevB.34.5390} {\bibfield  {journal} {\bibinfo
  {journal} {Phys. Rev. B}\ }\textbf {\bibinfo {volume} {34}},\ \bibinfo
  {pages} {5390} (\bibinfo {year} {1986})}\BibitemShut {NoStop}%
\bibitem [{\citenamefont {Godby}\ \emph {et~al.}(1988)\citenamefont {Godby},
  \citenamefont {Schl\"uter},\ and\ \citenamefont {Sham}}]{Godby1988}%
  \BibitemOpen
  \bibfield  {author} {\bibinfo {author} {\bibfnamefont {R.~W.}\ \bibnamefont
  {Godby}}, \bibinfo {author} {\bibfnamefont {M.}~\bibnamefont {Schl\"uter}},\
  and\ \bibinfo {author} {\bibfnamefont {L.~J.}\ \bibnamefont {Sham}},\
  }\bibfield  {title} {\bibinfo {title} {Self-energy operators and
  exchange-correlation potentials in semiconductors},\ }\href
  {https://doi.org/10.1103/PhysRevB.37.10159} {\bibfield  {journal} {\bibinfo
  {journal} {Phys. Rev. B}\ }\textbf {\bibinfo {volume} {37}},\ \bibinfo
  {pages} {10159} (\bibinfo {year} {1988})}\BibitemShut {NoStop}%
\bibitem [{\citenamefont {Albrecht}\ \emph {et~al.}(1998)\citenamefont
  {Albrecht}, \citenamefont {Reining}, \citenamefont {Del~Sole},\ and\
  \citenamefont {Onida}}]{Albrecht1998}%
  \BibitemOpen
  \bibfield  {author} {\bibinfo {author} {\bibfnamefont {S.}~\bibnamefont
  {Albrecht}}, \bibinfo {author} {\bibfnamefont {L.}~\bibnamefont {Reining}},
  \bibinfo {author} {\bibfnamefont {R.}~\bibnamefont {Del~Sole}},\ and\
  \bibinfo {author} {\bibfnamefont {G.}~\bibnamefont {Onida}},\ }\bibfield
  {title} {\bibinfo {title} {Ab initio calculation of excitonic effects in the
  optical spectra of semiconductors},\ }\href
  {https://doi.org/10.1103/PhysRevLett.80.4510} {\bibfield  {journal} {\bibinfo
   {journal} {Phys. Rev. Lett.}\ }\textbf {\bibinfo {volume} {80}},\ \bibinfo
  {pages} {4510} (\bibinfo {year} {1998})}\BibitemShut {NoStop}%
\bibitem [{\citenamefont {Benedict}\ \emph {et~al.}(1998)\citenamefont
  {Benedict}, \citenamefont {Shirley},\ and\ \citenamefont
  {Bohn}}]{Benedict1998}%
  \BibitemOpen
  \bibfield  {author} {\bibinfo {author} {\bibfnamefont {L.~X.}\ \bibnamefont
  {Benedict}}, \bibinfo {author} {\bibfnamefont {E.~L.}\ \bibnamefont
  {Shirley}},\ and\ \bibinfo {author} {\bibfnamefont {R.~B.}\ \bibnamefont
  {Bohn}},\ }\bibfield  {title} {\bibinfo {title} {Optical absorption of
  insulators and the electron-hole interaction: An ab initio calculation},\
  }\href {https://doi.org/10.1103/PhysRevLett.80.4514} {\bibfield  {journal}
  {\bibinfo  {journal} {Phys. Rev. Lett.}\ }\textbf {\bibinfo {volume} {80}},\
  \bibinfo {pages} {4514} (\bibinfo {year} {1998})}\BibitemShut {NoStop}%
\bibitem [{\citenamefont {Rohlfing}\ and\ \citenamefont
  {Louie}(2000)}]{Rohlfing2000}%
  \BibitemOpen
  \bibfield  {author} {\bibinfo {author} {\bibfnamefont {M.}~\bibnamefont
  {Rohlfing}}\ and\ \bibinfo {author} {\bibfnamefont {S.~G.}\ \bibnamefont
  {Louie}},\ }\bibfield  {title} {\bibinfo {title} {Electron-hole excitations
  and optical spectra from first principles},\ }\href
  {https://doi.org/10.1103/PhysRevB.62.4927} {\bibfield  {journal} {\bibinfo
  {journal} {Phys. Rev. B}\ }\textbf {\bibinfo {volume} {62}},\ \bibinfo
  {pages} {4927} (\bibinfo {year} {2000})}\BibitemShut {NoStop}%
\bibitem [{\citenamefont {Gonze}\ \emph {et~al.}(2005)\citenamefont {Gonze},
  \citenamefont {Rignanese}, \citenamefont {Verstraete}, \citenamefont
  {Beuken}, \citenamefont {Pouillon}, \citenamefont {Caracas}, \citenamefont
  {Jollet}, \citenamefont {Torrent}, \citenamefont {Zerah}, \citenamefont
  {Mikami} \emph {et~al.}}]{Gonze2005}%
  \BibitemOpen
  \bibfield  {author} {\bibinfo {author} {\bibfnamefont {X.}~\bibnamefont
  {Gonze}}, \bibinfo {author} {\bibfnamefont {G.-M.}\ \bibnamefont
  {Rignanese}}, \bibinfo {author} {\bibfnamefont {M.}~\bibnamefont
  {Verstraete}}, \bibinfo {author} {\bibfnamefont {J.-M.}\ \bibnamefont
  {Beuken}}, \bibinfo {author} {\bibfnamefont {Y.}~\bibnamefont {Pouillon}},
  \bibinfo {author} {\bibfnamefont {R.}~\bibnamefont {Caracas}}, \bibinfo
  {author} {\bibfnamefont {F.}~\bibnamefont {Jollet}}, \bibinfo {author}
  {\bibfnamefont {M.}~\bibnamefont {Torrent}}, \bibinfo {author} {\bibfnamefont
  {G.}~\bibnamefont {Zerah}}, \bibinfo {author} {\bibfnamefont
  {M.}~\bibnamefont {Mikami}}, \emph {et~al.},\ }\bibfield  {title} {\bibinfo
  {title} {A brief introduction to the abinit software package},\ }\href@noop
  {} {\bibfield  {journal} {\bibinfo  {journal} {Z. Kristallogr}\ }\textbf
  {\bibinfo {volume} {220}},\ \bibinfo {pages} {558} (\bibinfo {year}
  {2005})}\BibitemShut {NoStop}%
\bibitem [{EXC()}]{EXCcode}%
  \BibitemOpen
  \href@noop {} {}\bibinfo {note}
  {\url{http://www.bethe-salpeter.org/}}\BibitemShut {NoStop}%
\bibitem [{\citenamefont {H\"user}\ \emph {et~al.}(2013)\citenamefont
  {H\"user}, \citenamefont {Olsen},\ and\ \citenamefont
  {Thygesen}}]{Falco2013}%
  \BibitemOpen
  \bibfield  {author} {\bibinfo {author} {\bibfnamefont {F.}~\bibnamefont
  {H\"user}}, \bibinfo {author} {\bibfnamefont {T.}~\bibnamefont {Olsen}},\
  and\ \bibinfo {author} {\bibfnamefont {K.~S.}\ \bibnamefont {Thygesen}},\
  }\bibfield  {title} {\bibinfo {title} {Quasiparticle gw calculations for
  solids, molecules, and two-dimensional materials},\ }\href
  {https://doi.org/10.1103/PhysRevB.87.235132} {\bibfield  {journal} {\bibinfo
  {journal} {Phys. Rev. B}\ }\textbf {\bibinfo {volume} {87}},\ \bibinfo
  {pages} {235132} (\bibinfo {year} {2013})}\BibitemShut {NoStop}%
\bibitem [{\citenamefont {Gorelov}\ \emph
  {et~al.}(2023{\natexlab{a}})\citenamefont {Gorelov}, \citenamefont {Reining},
  \citenamefont {Lambrecht},\ and\ \citenamefont {Gatti}}]{Gorelov2023_v2o5}%
  \BibitemOpen
  \bibfield  {author} {\bibinfo {author} {\bibfnamefont {V.}~\bibnamefont
  {Gorelov}}, \bibinfo {author} {\bibfnamefont {L.}~\bibnamefont {Reining}},
  \bibinfo {author} {\bibfnamefont {W.~R.~L.}\ \bibnamefont {Lambrecht}},\ and\
  \bibinfo {author} {\bibfnamefont {M.}~\bibnamefont {Gatti}},\ }\bibfield
  {title} {\bibinfo {title} {{Robustness of electronic screening effects in
  electron spectroscopies: Example of V$_2$O$_5$}},\ }\href
  {https://doi.org/10.1103/PhysRevB.107.075101} {\bibfield  {journal} {\bibinfo
   {journal} {Physical Review B}\ }\textbf {\bibinfo {volume} {107}},\ \bibinfo
  {pages} {075101} (\bibinfo {year} {2023}{\natexlab{a}})}\BibitemShut
  {NoStop}%
\bibitem [{\citenamefont {Gorelov}\ \emph {et~al.}(2022)\citenamefont
  {Gorelov}, \citenamefont {Reining}, \citenamefont {Feneberg}, \citenamefont
  {Goldhahn}, \citenamefont {Schleife}, \citenamefont {Lambrecht},\ and\
  \citenamefont {Gatti}}]{Gorelov2022}%
  \BibitemOpen
  \bibfield  {author} {\bibinfo {author} {\bibfnamefont {V.}~\bibnamefont
  {Gorelov}}, \bibinfo {author} {\bibfnamefont {L.}~\bibnamefont {Reining}},
  \bibinfo {author} {\bibfnamefont {M.}~\bibnamefont {Feneberg}}, \bibinfo
  {author} {\bibfnamefont {R.}~\bibnamefont {Goldhahn}}, \bibinfo {author}
  {\bibfnamefont {A.}~\bibnamefont {Schleife}}, \bibinfo {author}
  {\bibfnamefont {W.~R.~L.}\ \bibnamefont {Lambrecht}},\ and\ \bibinfo {author}
  {\bibfnamefont {M.}~\bibnamefont {Gatti}},\ }\bibfield  {title} {\bibinfo
  {title} {{Delocalization of dark and bright excitons in flat-band materials
  and the optical properties of V2O5}},\ }\href
  {https://doi.org/10.1038/s41524-022-00754-2} {\bibfield  {journal} {\bibinfo
  {journal} {npj Computational Materials}\ }\textbf {\bibinfo {volume} {8}},\
  \bibinfo {pages} {94} (\bibinfo {year} {2022})}\BibitemShut {NoStop}%
\bibitem [{\citenamefont {Gorelov}\ \emph
  {et~al.}(2020{\natexlab{c}})\citenamefont {Gorelov}, \citenamefont
  {Ceperley}, \citenamefont {Holzmann},\ and\ \citenamefont
  {Pierleoni}}]{Gorelov2020b}%
  \BibitemOpen
  \bibfield  {author} {\bibinfo {author} {\bibfnamefont {V.}~\bibnamefont
  {Gorelov}}, \bibinfo {author} {\bibfnamefont {D.~M.}\ \bibnamefont
  {Ceperley}}, \bibinfo {author} {\bibfnamefont {M.}~\bibnamefont {Holzmann}},\
  and\ \bibinfo {author} {\bibfnamefont {C.}~\bibnamefont {Pierleoni}},\
  }\bibfield  {title} {\bibinfo {title} {{Electronic structure and optical
  properties of quantum crystals from first principles calculations in the
  Born-Oppenheimer approximation}},\ }\href {https://doi.org/10.1063/5.0031843}
  {\bibfield  {journal} {\bibinfo  {journal} {J. Chem Phys}\ }\textbf {\bibinfo
  {volume} {153}},\ \bibinfo {pages} {234117} (\bibinfo {year}
  {2020}{\natexlab{c}})},\ \Eprint {https://arxiv.org/abs/2010.01988}
  {arXiv:2010.01988} \BibitemShut {NoStop}%
\bibitem [{\citenamefont {Makov}\ and\ \citenamefont
  {Payne}(1995)}]{Makov1995}%
  \BibitemOpen
  \bibfield  {author} {\bibinfo {author} {\bibfnamefont {G.}~\bibnamefont
  {Makov}}\ and\ \bibinfo {author} {\bibfnamefont {M.~C.}\ \bibnamefont
  {Payne}},\ }\bibfield  {title} {\bibinfo {title} {Periodic boundary
  conditions in ab initio calculations},\ }\href
  {https://doi.org/10.1103/PhysRevB.51.4014} {\bibfield  {journal} {\bibinfo
  {journal} {Phys. Rev. B}\ }\textbf {\bibinfo {volume} {51}},\ \bibinfo
  {pages} {4014} (\bibinfo {year} {1995})}\BibitemShut {NoStop}%
\bibitem [{\citenamefont {Engel}\ \emph {et~al.}(1995)\citenamefont {Engel},
  \citenamefont {Kwon},\ and\ \citenamefont {Martin}}]{Engel1995}%
  \BibitemOpen
  \bibfield  {author} {\bibinfo {author} {\bibfnamefont {G.~E.}\ \bibnamefont
  {Engel}}, \bibinfo {author} {\bibfnamefont {Y.}~\bibnamefont {Kwon}},\ and\
  \bibinfo {author} {\bibfnamefont {R.~M.}\ \bibnamefont {Martin}},\ }\bibfield
   {title} {\bibinfo {title} {Quasiparticle bands in a two-dimensional crystal
  found by gw and quantum monte carlo calculations},\ }\href
  {https://doi.org/10.1103/PhysRevB.51.13538} {\bibfield  {journal} {\bibinfo
  {journal} {Phys. Rev. B}\ }\textbf {\bibinfo {volume} {51}},\ \bibinfo
  {pages} {13538} (\bibinfo {year} {1995})}\BibitemShut {NoStop}%
\bibitem [{\citenamefont {Gorelov}\ \emph
  {et~al.}(2023{\natexlab{b}})\citenamefont {Gorelov}, \citenamefont {Yang},
  \citenamefont {Ruggeri}, \citenamefont {Ceperley}, \citenamefont
  {Pierleoni},\ and\ \citenamefont {Holzmann}}]{Gorelov2023}%
  \BibitemOpen
  \bibfield  {author} {\bibinfo {author} {\bibfnamefont {V.}~\bibnamefont
  {Gorelov}}, \bibinfo {author} {\bibfnamefont {Y.}~\bibnamefont {Yang}},
  \bibinfo {author} {\bibfnamefont {M.}~\bibnamefont {Ruggeri}}, \bibinfo
  {author} {\bibfnamefont {D.~M.}\ \bibnamefont {Ceperley}}, \bibinfo {author}
  {\bibfnamefont {C.}~\bibnamefont {Pierleoni}},\ and\ \bibinfo {author}
  {\bibfnamefont {M.}~\bibnamefont {Holzmann}},\ }\bibfield  {title} {\bibinfo
  {title} {Neutral band gap of carbon by quantum monte carlo methods},\ }\href
  {https://doi.org/10.5488/CMP.26.33701} {\bibfield  {journal} {\bibinfo
  {journal} {Condensed Matter Physics}\ }\textbf {\bibinfo {volume} {26}},\
  \bibinfo {pages} {33701} (\bibinfo {year} {2023}{\natexlab{b}})}\BibitemShut
  {NoStop}%
\bibitem [{\citenamefont {Hunt}\ \emph {et~al.}(2018)\citenamefont {Hunt},
  \citenamefont {Szyniszewski}, \citenamefont {Prayogo}, \citenamefont
  {Maezono},\ and\ \citenamefont {Drummond}}]{Hunt2018}%
  \BibitemOpen
  \bibfield  {author} {\bibinfo {author} {\bibfnamefont {R.~J.}\ \bibnamefont
  {Hunt}}, \bibinfo {author} {\bibfnamefont {M.}~\bibnamefont {Szyniszewski}},
  \bibinfo {author} {\bibfnamefont {G.~I.}\ \bibnamefont {Prayogo}}, \bibinfo
  {author} {\bibfnamefont {R.}~\bibnamefont {Maezono}},\ and\ \bibinfo {author}
  {\bibfnamefont {N.~D.}\ \bibnamefont {Drummond}},\ }\bibfield  {title}
  {\bibinfo {title} {{Quantum Monte Carlo calculations of energy gaps from
  first principles}},\ }\href {https://doi.org/10.1103/PhysRevB.98.075122}
  {\bibfield  {journal} {\bibinfo  {journal} {Physical Review B}\ }\textbf
  {\bibinfo {volume} {98}},\ \bibinfo {pages} {1} (\bibinfo {year} {2018})},\
  \Eprint {https://arxiv.org/abs/1806.04750} {arXiv:1806.04750} \BibitemShut
  {NoStop}%
\bibitem [{\citenamefont {Cudazzo}\ \emph {et~al.}(2013)\citenamefont
  {Cudazzo}, \citenamefont {Gatti}, \citenamefont {Rubio},\ and\ \citenamefont
  {Sottile}}]{Cudazzo2013}%
  \BibitemOpen
  \bibfield  {author} {\bibinfo {author} {\bibfnamefont {P.}~\bibnamefont
  {Cudazzo}}, \bibinfo {author} {\bibfnamefont {M.}~\bibnamefont {Gatti}},
  \bibinfo {author} {\bibfnamefont {A.}~\bibnamefont {Rubio}},\ and\ \bibinfo
  {author} {\bibfnamefont {F.}~\bibnamefont {Sottile}},\ }\bibfield  {title}
  {\bibinfo {title} {{Frenkel versus charge-transfer exciton dispersion in
  molecular crystals}},\ }\href {https://doi.org/10.1103/PhysRevB.88.195152}
  {\bibfield  {journal} {\bibinfo  {journal} {Physical Review B - Condensed
  Matter and Materials Physics}\ }\textbf {\bibinfo {volume} {88}},\ \bibinfo
  {pages} {195152} (\bibinfo {year} {2013})}\BibitemShut {NoStop}%
\bibitem [{\citenamefont {Inoue}\ \emph {et~al.}(1979)\citenamefont {Inoue},
  \citenamefont {Kanzaki},\ and\ \citenamefont {Suga}}]{Inoue1979}%
  \BibitemOpen
  \bibfield  {author} {\bibinfo {author} {\bibfnamefont {K.}~\bibnamefont
  {Inoue}}, \bibinfo {author} {\bibfnamefont {H.}~\bibnamefont {Kanzaki}},\
  and\ \bibinfo {author} {\bibfnamefont {S.}~\bibnamefont {Suga}},\ }\bibfield
  {title} {\bibinfo {title} {Fundamental absorption spectra of solid
  hydrogen},\ }\href {https://doi.org/10.1016/0038-1098(79)90110-8} {\bibfield
  {journal} {\bibinfo  {journal} {Solid State Communications}\ }\textbf
  {\bibinfo {volume} {30}},\ \bibinfo {pages} {627} (\bibinfo {year}
  {1979})}\BibitemShut {NoStop}%
\bibitem [{\citenamefont {Loubeyre}\ \emph {et~al.}(2002)\citenamefont
  {Loubeyre}, \citenamefont {Occelli},\ and\ \citenamefont
  {LeToullec}}]{Loubeyre2002}%
  \BibitemOpen
  \bibfield  {author} {\bibinfo {author} {\bibfnamefont {P.}~\bibnamefont
  {Loubeyre}}, \bibinfo {author} {\bibfnamefont {F.}~\bibnamefont {Occelli}},\
  and\ \bibinfo {author} {\bibfnamefont {R.}~\bibnamefont {LeToullec}},\
  }\bibfield  {title} {\bibinfo {title} {{Optical studies of solid hydrogen to
  320 GPa and evidence for black hydrogen}},\ }\href
  {papers://68dd8153-82a1-4a0e-9ae0-80ae65a14bf7/Paper/p701} {\bibfield
  {journal} {\bibinfo  {journal} {Nature}\ }\textbf {\bibinfo {volume} {416}},\
  \bibinfo {pages} {613} (\bibinfo {year} {2002})}\BibitemShut {NoStop}%
\bibitem [{\citenamefont {Beutler}\ and\ \citenamefont
  {Junger}(1936)}]{Beutler1936}%
  \BibitemOpen
  \bibfield  {author} {\bibinfo {author} {\bibfnamefont {H.}~\bibnamefont
  {Beutler}}\ and\ \bibinfo {author} {\bibfnamefont {H.~O.}\ \bibnamefont
  {Junger}},\ }\bibfield  {title} {\bibinfo {title} {{Uber das
  Absorptionsspektrum des Wasserstoffs. III}},\ }\href
  {https://doi.org/10.1007/BF01337746} {\bibfield  {journal} {\bibinfo
  {journal} {Zeitschrift fur Physik}\ }\textbf {\bibinfo {volume} {100}},\
  \bibinfo {pages} {80} (\bibinfo {year} {1936})}\BibitemShut {NoStop}%
\bibitem [{\citenamefont {Herzberg}(1950)}]{Herzberg1950}%
  \BibitemOpen
  \bibfield  {author} {\bibinfo {author} {\bibfnamefont {G.}~\bibnamefont
  {Herzberg}},\ }\href@noop {} {\emph {\bibinfo {title} {{Molecular Spectra and
  Molecular Structure. Volume I: Spectra of Diatomic Molecules}}}},\ \bibinfo
  {edition} {2nd}\ ed.\ (\bibinfo  {publisher} {D. Van Nostrand},\ \bibinfo
  {address} {New York, NY},\ \bibinfo {year} {1950})\ p.\ \bibinfo {pages}
  {658}\BibitemShut {NoStop}%
\bibitem [{\citenamefont {Davydov}(1971)}]{Davydov1971}%
  \BibitemOpen
  \bibfield  {author} {\bibinfo {author} {\bibfnamefont {A.~S.}\ \bibnamefont
  {Davydov}},\ }\href {https://doi.org/10.1007/978-1-4899-5169-4} {\emph
  {\bibinfo {title} {{Theory of Molecular Excitons}}}},\ \bibinfo {edition}
  {1st}\ ed.\ (\bibinfo  {publisher} {Springer New York, NY},\ \bibinfo
  {address} {New York, NY},\ \bibinfo {year} {1971})\ p.\ \bibinfo {pages}
  {313}\BibitemShut {NoStop}%
\bibitem [{Note1()}]{Note1}%
  \BibitemOpen
  \bibinfo {note} {We have verified that the first exciton energies in
  absorption spectra are identical to those in the energy loss
  spectra}\BibitemShut {NoStop}%
\end{thebibliography}%

\end{document}